\documentclass[a4paper,UKenglish,cleveref, autoref, thm-restate,authorcolumns]{lipics-v2019}
\usepackage{cite}
\usepackage{booktabs}
\usepackage{pifont}%
\usepackage{rotating}
\usepackage{amsfonts}
\usepackage{multirow}
\usepackage{color}
\usepackage{graphicx}
\usepackage{hyperref}
\usepackage{amsmath}
\usepackage{algpseudocode}
\usepackage{algorithmicx}
\usepackage{algorithm} %
\usepackage{flushend}
\usepackage{subcaption}
\usepackage{wrapfig}
\usepackage{breakurl}
\usepackage{enumitem}
\usepackage{xspace}
\usepackage[normalem]{ulem}
\usepackage{listings}
\usepackage{flushend}
\usepackage{footmisc}

\makeatletter
\newcommand{\@chapapp}{\relax}%
\makeatother
\usepackage[title,toc,titletoc,header]{appendix}

\usepackage{tikz}
\usetikzlibrary{positioning, shapes}

\usepackage[version=4]{mhchem} %
\usepackage{float} %

\usepackage{lipsum}

\newcommand{\DefMacro}[2]{\expandafter\newcommand\csname rmk-#1\endcsname{#2}}
\newcommand{\UseMacro}[1]{\csname rmk-#1\endcsname}

\newcommand{\CodeIn}[1]{\begin{small}\texttt{#1}\end{small}}
\newcommand{\XComment}[1]{}
\newcommand{\Space}[1]{}

\newcommand{\TNC}[2]{\multicolumn{#1}{c}{#2}}

\definecolor{gray}{RGB}{211,211,211}
\newcommand{\jbasicstyle}{\small\sffamily}

\newcommand{\jnumberstyle}{\scriptsize}

\lstdefinelanguage{pseudo}
{ morekeywords={for, in, break, continue, try, except, not,
  if,else,return,map,fieldElement_array_array40,fieldElement_array40},
  keywordstyle=\bfseries, lineskip=-0.1em, numbers=left,
  numberstyle=\jnumberstyle, numbersep=4pt, basicstyle=\jbasicstyle,
  breaklines=true, breakautoindent=true, tabsize=2,
  columns=fullflexible, morecomment=*[l][\textsl]{//},
  mathescape=true, }
\newenvironment{CodeOut}{\begin{scriptsize}}{\end{scriptsize}}

\def\formulaname{Formula}

\newfloat{formula}{pb}{form}
\floatname{formula}{\formulaname}

\newcommand{\Relu}{\textit{ReLU}\xspace}

\newcommand{\Section}[1]{\vspace*{-2ex}\section{#1}\vspace*{-2ex}}
\newcommand{\Caption}[1]{\vspace*{-2ex}\caption{#1}\vspace*{-1ex}}

\newcommand{\BeginTable}{\begin{small} \begin{table}[!t]}
\newcommand{\EndTable}{\end{table} \end{small}}

\usepackage[textwidth=60pt,textsize=scriptsize]{todonotes} 
\LetLtxMacro{\todom}{\todo}
\renewcommand{\todo}[1]{\todom[inline]{#1}}

\algnewcommand\algorithmicforeach{\textbf{for each}}
\algdef{S}[FOR]{ForEach}[1]{\algorithmicforeach\ #1\ \algorithmicdo}

\algnewcommand{\IfThen}[2]{%
  \State \algorithmicif\ #1\ \algorithmicthen\ #2}

\DefMacro{seesaw_D4_R2_S20_crns}{9}
\DefMacro{seesaw_D4_R2_S20_time}{00:00:01}
\DefMacro{seesaw_D4_R4_S20_crns}{92}
\DefMacro{seesaw_D4_R4_S20_time}{00:00:18}
\DefMacro{seesaw_D4_R2_S15_crns}{9}
\DefMacro{seesaw_D4_R2_S15_time}{00:00:07}
\DefMacro{seesaw_D5_R4_S12_crns}{136}
\DefMacro{seesaw_D5_R4_S12_time}{00:00:31}
\DefMacro{seesaw_D4_R3_S15_crns}{33}
\DefMacro{seesaw_D4_R3_S15_time}{00:02:32}
\DefMacro{seesaw_D3_R3_S10_crns}{15}
\DefMacro{seesaw_D3_R3_S10_time}{00:00:20}
\DefMacro{seesaw_D1_R2_S8_crns}{0}
\DefMacro{seesaw_D1_R2_S8_time}{00:00:00}
\DefMacro{seesaw_D3_R1_S10_crns}{1}
\DefMacro{seesaw_D3_R1_S10_time}{00:00:00}
\DefMacro{seesaw_D3_R3_S20_crns}{15}
\DefMacro{seesaw_D3_R3_S20_time}{00:00:01}
\DefMacro{seesaw_D2_R5_S8_crns}{0}
\DefMacro{seesaw_D2_R5_S8_time}{00:00:04}
\DefMacro{seesaw_D5_R2_S20_crns}{4}
\DefMacro{seesaw_D5_R2_S20_time}{00:00:00}
\DefMacro{seesaw_D3_R4_S8_crns}{2}
\DefMacro{seesaw_D3_R4_S8_time}{00:00:15}
\DefMacro{seesaw_D1_R4_S20_crns}{0}
\DefMacro{seesaw_D1_R4_S20_time}{00:00:00}
\DefMacro{seesaw_D4_R4_S12_crns}{88}
\DefMacro{seesaw_D4_R4_S12_time}{00:00:19}
\DefMacro{seesaw_D7_R4_S20_crns}{424}
\DefMacro{seesaw_D7_R4_S20_time}{03:46:59}
\DefMacro{seesaw_D1_R1_S8_crns}{1}
\DefMacro{seesaw_D1_R1_S8_time}{00:00:00}
\DefMacro{seesaw_D4_R5_S10_crns}{4}
\DefMacro{seesaw_D4_R5_S10_time}{00:01:11}
\DefMacro{seesaw_D3_R3_S12_crns}{15}
\DefMacro{seesaw_D3_R3_S12_time}{00:00:02}
\DefMacro{seesaw_D2_R1_S8_crns}{1}
\DefMacro{seesaw_D2_R1_S8_time}{00:00:01}
\DefMacro{seesaw_D1_R2_S15_crns}{0}
\DefMacro{seesaw_D1_R2_S15_time}{00:00:00}
\DefMacro{seesaw_D4_R1_S15_crns}{0}
\DefMacro{seesaw_D4_R1_S15_time}{00:00:00}
\DefMacro{seesaw_D5_R3_S20_crns}{55}
\DefMacro{seesaw_D5_R3_S20_time}{00:00:04}
\DefMacro{seesaw_D2_R4_S15_crns}{2}
\DefMacro{seesaw_D2_R4_S15_time}{00:00:09}
\DefMacro{seesaw_D3_R2_S8_crns}{5}
\DefMacro{seesaw_D3_R2_S8_time}{00:00:05}
\DefMacro{seesaw_D1_R3_S10_crns}{0}
\DefMacro{seesaw_D1_R3_S10_time}{00:00:00}
\DefMacro{seesaw_D1_R5_S8_crns}{0}
\DefMacro{seesaw_D1_R5_S8_time}{00:00:01}
\DefMacro{seesaw_D5_R1_S10_crns}{0}
\DefMacro{seesaw_D5_R1_S10_time}{00:00:00}
\DefMacro{seesaw_D3_R5_S10_crns}{4}
\DefMacro{seesaw_D3_R5_S10_time}{00:00:48}
\DefMacro{seesaw_D3_R3_S8_crns}{8}
\DefMacro{seesaw_D3_R3_S8_time}{00:00:11}
\DefMacro{seesaw_D7_R2_S20_crns}{0}
\DefMacro{seesaw_D7_R2_S20_time}{00:00:00}
\DefMacro{seesaw_D3_R4_S15_crns}{13}
\DefMacro{seesaw_D3_R4_S15_time}{00:01:21}
\DefMacro{seesaw_D3_R5_S20_crns}{14}
\DefMacro{seesaw_D3_R5_S20_time}{00:00:17}
\DefMacro{seesaw_D1_R3_S15_crns}{0}
\DefMacro{seesaw_D1_R3_S15_time}{00:00:01}
\DefMacro{seesaw_D1_R4_S8_crns}{0}
\DefMacro{seesaw_D1_R4_S8_time}{00:00:01}
\DefMacro{seesaw_D2_R2_S20_crns}{4}
\DefMacro{seesaw_D2_R2_S20_time}{00:00:00}
\DefMacro{seesaw_D7_R3_S20_crns}{20}
\DefMacro{seesaw_D7_R3_S20_time}{00:00:33}
\DefMacro{seesaw_D1_R5_S15_crns}{0}
\DefMacro{seesaw_D1_R5_S15_time}{00:00:01}
\DefMacro{seesaw_D3_R2_S20_crns}{5}
\DefMacro{seesaw_D3_R2_S20_time}{00:00:00}
\DefMacro{seesaw_D2_R5_S15_crns}{1}
\DefMacro{seesaw_D2_R5_S15_time}{00:00:11}
\DefMacro{seesaw_D5_R4_S20_crns}{243}
\DefMacro{seesaw_D5_R4_S20_time}{00:00:48}
\DefMacro{seesaw_D5_R5_S12_crns}{49}
\DefMacro{seesaw_D5_R5_S12_time}{00:00:39}
\DefMacro{seesaw_D1_R4_S12_crns}{0}
\DefMacro{seesaw_D1_R4_S12_time}{00:00:00}
\DefMacro{seesaw_D4_R1_S10_crns}{0}
\DefMacro{seesaw_D4_R1_S10_time}{00:00:00}
\DefMacro{seesaw_D2_R2_S12_crns}{4}
\DefMacro{seesaw_D2_R2_S12_time}{00:00:00}
\DefMacro{seesaw_D2_R2_S10_crns}{4}
\DefMacro{seesaw_D2_R2_S10_time}{00:00:01}
\DefMacro{seesaw_D3_R1_S8_crns}{1}
\DefMacro{seesaw_D3_R1_S8_time}{00:00:01}
\DefMacro{seesaw_D2_R3_S8_crns}{0}
\DefMacro{seesaw_D2_R3_S8_time}{00:00:01}
\DefMacro{seesaw_D5_R2_S12_crns}{4}
\DefMacro{seesaw_D5_R2_S12_time}{00:00:00}
\DefMacro{seesaw_D1_R2_S20_crns}{0}
\DefMacro{seesaw_D1_R2_S20_time}{00:00:00}
\DefMacro{seesaw_D2_R1_S10_crns}{1}
\DefMacro{seesaw_D2_R1_S10_time}{00:00:00}
\DefMacro{seesaw_D4_R2_S10_crns}{9}
\DefMacro{seesaw_D4_R2_S10_time}{00:00:04}
\DefMacro{seesaw_D2_R4_S12_crns}{2}
\DefMacro{seesaw_D2_R4_S12_time}{00:00:01}
\DefMacro{seesaw_D1_R3_S12_crns}{0}
\DefMacro{seesaw_D1_R3_S12_time}{00:00:00}
\DefMacro{seesaw_D5_R4_S10_crns}{3}
\DefMacro{seesaw_D5_R4_S10_time}{00:00:20}
\DefMacro{seesaw_D3_R4_S10_crns}{12}
\DefMacro{seesaw_D3_R4_S10_time}{00:00:40}
\DefMacro{seesaw_D3_R2_S12_crns}{5}
\DefMacro{seesaw_D3_R2_S12_time}{00:00:00}
\DefMacro{seesaw_D1_R1_S12_crns}{1}
\DefMacro{seesaw_D1_R1_S12_time}{00:00:00}
\DefMacro{seesaw_D3_R1_S20_crns}{1}
\DefMacro{seesaw_D3_R1_S20_time}{00:00:00}
\DefMacro{seesaw_D3_R1_S15_crns}{1}
\DefMacro{seesaw_D3_R1_S15_time}{00:00:00}
\DefMacro{seesaw_D6_R1_S20_crns}{0}
\DefMacro{seesaw_D6_R1_S20_time}{00:00:00}
\DefMacro{seesaw_D3_R4_S12_crns}{13}
\DefMacro{seesaw_D3_R4_S12_time}{00:00:09}
\DefMacro{seesaw_D4_R4_S10_crns}{19}
\DefMacro{seesaw_D4_R4_S10_time}{00:02:02}
\DefMacro{seesaw_D4_R4_S15_crns}{92}
\DefMacro{seesaw_D4_R4_S15_time}{02:22:45}
\DefMacro{seesaw_D1_R1_S15_crns}{1}
\DefMacro{seesaw_D1_R1_S15_time}{00:00:00}
\DefMacro{seesaw_D2_R4_S8_crns}{1}
\DefMacro{seesaw_D2_R4_S8_time}{00:00:09}
\DefMacro{seesaw_D5_R5_S20_crns}{705}
\DefMacro{seesaw_D5_R5_S20_time}{00:10:16}
\DefMacro{seesaw_D3_R1_S12_crns}{1}
\DefMacro{seesaw_D3_R1_S12_time}{00:00:00}
\DefMacro{seesaw_D4_R3_S20_crns}{33}
\DefMacro{seesaw_D4_R3_S20_time}{00:00:02}
\DefMacro{seesaw_D3_R3_S15_crns}{15}
\DefMacro{seesaw_D3_R3_S15_time}{00:00:24}
\DefMacro{seesaw_D2_R1_S15_crns}{1}
\DefMacro{seesaw_D2_R1_S15_time}{00:00:01}
\DefMacro{seesaw_D8_R4_S20_crns}{ERROR}
\DefMacro{seesaw_D8_R4_S20_time}{ERROR}
\DefMacro{seesaw_D2_R3_S15_crns}{0}
\DefMacro{seesaw_D2_R3_S15_time}{00:00:04}
\DefMacro{seesaw_D2_R1_S20_crns}{1}
\DefMacro{seesaw_D2_R1_S20_time}{00:00:00}
\DefMacro{seesaw_D8_R3_S20_crns}{4}
\DefMacro{seesaw_D8_R3_S20_time}{00:01:13}
\DefMacro{seesaw_D6_R4_S20_crns}{436}
\DefMacro{seesaw_D6_R4_S20_time}{00:06:40}
\DefMacro{seesaw_D5_R5_S10_crns}{0}
\DefMacro{seesaw_D5_R5_S10_time}{00:00:29}
\DefMacro{seesaw_D1_R3_S20_crns}{0}
\DefMacro{seesaw_D1_R3_S20_time}{00:00:00}
\DefMacro{seesaw_D3_R5_S8_crns}{0}
\DefMacro{seesaw_D3_R5_S8_time}{00:00:14}
\DefMacro{seesaw_D4_R5_S12_crns}{38}
\DefMacro{seesaw_D4_R5_S12_time}{00:00:48}
\DefMacro{seesaw_D3_R5_S15_crns}{14}
\DefMacro{seesaw_D3_R5_S15_time}{00:08:13}
\DefMacro{seesaw_D4_R1_S20_crns}{0}
\DefMacro{seesaw_D4_R1_S20_time}{00:00:00}
\DefMacro{seesaw_D6_R5_S20_crns}{2027}
\DefMacro{seesaw_D6_R5_S20_time}{03:01:06}
\DefMacro{seesaw_D4_R3_S12_crns}{33}
\DefMacro{seesaw_D4_R3_S12_time}{00:00:04}
\DefMacro{seesaw_D2_R1_S12_crns}{1}
\DefMacro{seesaw_D2_R1_S12_time}{00:00:00}
\DefMacro{seesaw_D5_R1_S12_crns}{0}
\DefMacro{seesaw_D5_R1_S12_time}{00:00:00}
\DefMacro{seesaw_D1_R2_S12_crns}{0}
\DefMacro{seesaw_D1_R2_S12_time}{00:00:00}
\DefMacro{seesaw_D2_R5_S12_crns}{1}
\DefMacro{seesaw_D2_R5_S12_time}{00:00:02}
\DefMacro{seesaw_D5_R3_S12_crns}{55}
\DefMacro{seesaw_D5_R3_S12_time}{00:00:06}
\DefMacro{seesaw_D7_R1_S20_crns}{0}
\DefMacro{seesaw_D7_R1_S20_time}{00:00:00}
\DefMacro{seesaw_D7_R5_S20_crns}{ERROR}
\DefMacro{seesaw_D7_R5_S20_time}{ERROR}
\DefMacro{seesaw_D4_R1_S12_crns}{0}
\DefMacro{seesaw_D4_R1_S12_time}{00:00:00}
\DefMacro{seesaw_D2_R4_S10_crns}{2}
\DefMacro{seesaw_D2_R4_S10_time}{00:00:04}
\DefMacro{seesaw_D4_R3_S10_crns}{30}
\DefMacro{seesaw_D4_R3_S10_time}{00:01:19}
\DefMacro{seesaw_D2_R4_S20_crns}{2}
\DefMacro{seesaw_D2_R4_S20_time}{00:00:01}
\DefMacro{seesaw_D5_R2_S10_crns}{4}
\DefMacro{seesaw_D5_R2_S10_time}{00:00:04}
\DefMacro{seesaw_D1_R5_S20_crns}{0}
\DefMacro{seesaw_D1_R5_S20_time}{00:00:00}
\DefMacro{seesaw_D3_R5_S12_crns}{14}
\DefMacro{seesaw_D3_R5_S12_time}{00:00:21}
\DefMacro{seesaw_D1_R5_S10_crns}{0}
\DefMacro{seesaw_D1_R5_S10_time}{00:00:00}
\DefMacro{seesaw_D2_R2_S8_crns}{4}
\DefMacro{seesaw_D2_R2_S8_time}{00:00:04}
\DefMacro{seesaw_D8_R5_S20_crns}{ERROR}
\DefMacro{seesaw_D8_R5_S20_time}{ERROR}
\DefMacro{seesaw_D2_R3_S20_crns}{0}
\DefMacro{seesaw_D2_R3_S20_time}{00:00:00}
\DefMacro{seesaw_D4_R2_S12_crns}{9}
\DefMacro{seesaw_D4_R2_S12_time}{00:00:00}
\DefMacro{seesaw_D2_R5_S20_crns}{1}
\DefMacro{seesaw_D2_R5_S20_time}{00:00:03}
\DefMacro{seesaw_D2_R5_S10_crns}{1}
\DefMacro{seesaw_D2_R5_S10_time}{00:00:07}
\DefMacro{seesaw_D3_R4_S20_crns}{13}
\DefMacro{seesaw_D3_R4_S20_time}{00:00:05}
\DefMacro{seesaw_D8_R1_S20_crns}{4}
\DefMacro{seesaw_D8_R1_S20_time}{00:01:13}
\DefMacro{seesaw_D8_R2_S20_crns}{4}
\DefMacro{seesaw_D8_R2_S20_time}{00:01:12}
\DefMacro{seesaw_D6_R2_S20_crns}{1}
\DefMacro{seesaw_D6_R2_S20_time}{00:00:00}
\DefMacro{seesaw_D1_R5_S12_crns}{0}
\DefMacro{seesaw_D1_R5_S12_time}{00:00:00}
\DefMacro{seesaw_D3_R2_S15_crns}{5}
\DefMacro{seesaw_D3_R2_S15_time}{00:00:03}
\DefMacro{seesaw_D4_R5_S20_crns}{121}
\DefMacro{seesaw_D4_R5_S20_time}{00:01:58}
\DefMacro{seesaw_D1_R3_S8_crns}{0}
\DefMacro{seesaw_D1_R3_S8_time}{00:00:01}
\DefMacro{seesaw_D2_R3_S10_crns}{0}
\DefMacro{seesaw_D2_R3_S10_time}{00:00:02}
\DefMacro{seesaw_D5_R1_S20_crns}{0}
\DefMacro{seesaw_D5_R1_S20_time}{00:00:00}
\DefMacro{seesaw_D1_R4_S15_crns}{0}
\DefMacro{seesaw_D1_R4_S15_time}{00:00:01}
\DefMacro{seesaw_D2_R2_S15_crns}{4}
\DefMacro{seesaw_D2_R2_S15_time}{00:00:03}
\DefMacro{seesaw_D5_R3_S10_crns}{34}
\DefMacro{seesaw_D5_R3_S10_time}{00:02:42}
\DefMacro{seesaw_D1_R2_S10_crns}{0}
\DefMacro{seesaw_D1_R2_S10_time}{00:00:00}
\DefMacro{seesaw_D1_R4_S10_crns}{0}
\DefMacro{seesaw_D1_R4_S10_time}{00:00:00}
\DefMacro{seesaw_D6_R3_S20_crns}{43}
\DefMacro{seesaw_D6_R3_S20_time}{00:00:10}
\DefMacro{seesaw_D2_R3_S12_crns}{0}
\DefMacro{seesaw_D2_R3_S12_time}{00:00:01}
\DefMacro{seesaw_D3_R2_S10_crns}{5}
\DefMacro{seesaw_D3_R2_S10_time}{00:00:02}
\DefMacro{seesaw_D1_R1_S10_crns}{1}
\DefMacro{seesaw_D1_R1_S10_time}{00:00:00}
\DefMacro{seesaw_D1_R1_S20_crns}{1}
\DefMacro{seesaw_D1_R1_S20_time}{00:00:00}
\DefMacro{seesaw-old_D1_R6_S8_crns}{0}
\DefMacro{seesaw-old_D1_R6_S8_time}{00:00:01}
\DefMacro{seesaw-old_D1_R6_S6_crns}{0}
\DefMacro{seesaw-old_D1_R6_S6_time}{00:00:02}
\DefMacro{seesaw-old_D3_R4_S6_crns}{2}
\DefMacro{seesaw-old_D3_R4_S6_time}{00:05:36}
\DefMacro{seesaw-old_D2_R2_S7_crns}{0}
\DefMacro{seesaw-old_D2_R2_S7_time}{00:00:01}
\DefMacro{seesaw-old_D2_R4_S2_crns}{0}
\DefMacro{seesaw-old_D2_R4_S2_time}{00:00:00}
\DefMacro{seesaw-old_D1_R2_S8_crns}{0}
\DefMacro{seesaw-old_D1_R2_S8_time}{00:00:04}
\DefMacro{seesaw-old_D3_R6_S6_crns}{0}
\DefMacro{seesaw-old_D3_R6_S6_time}{00:01:21}
\DefMacro{seesaw-old_D1_R2_S7_crns}{0}
\DefMacro{seesaw-old_D1_R2_S7_time}{00:00:01}
\DefMacro{seesaw-old_D3_R4_S8_crns}{1}
\DefMacro{seesaw-old_D3_R4_S8_time}{08:32:34}
\DefMacro{seesaw-old_D2_R6_S7_crns}{0}
\DefMacro{seesaw-old_D2_R6_S7_time}{00:03:08}
\DefMacro{seesaw-old_D2_R2_S2_crns}{0}
\DefMacro{seesaw-old_D2_R2_S2_time}{00:00:00}
\DefMacro{seesaw-old_D2_R2_S6_crns}{0}
\DefMacro{seesaw-old_D2_R2_S6_time}{00:00:01}
\DefMacro{seesaw-old_D1_R6_S2_crns}{0}
\DefMacro{seesaw-old_D1_R6_S2_time}{00:00:00}
\DefMacro{seesaw-old_D3_R2_S8_crns}{0}
\DefMacro{seesaw-old_D3_R2_S8_time}{00:00:01}
\DefMacro{seesaw-old_D2_R2_S3_crns}{0}
\DefMacro{seesaw-old_D2_R2_S3_time}{00:00:01}
\DefMacro{seesaw-old_D1_R6_S5_crns}{0}
\DefMacro{seesaw-old_D1_R6_S5_time}{00:00:01}
\DefMacro{seesaw-old_D3_R2_S7_crns}{0}
\DefMacro{seesaw-old_D3_R2_S7_time}{00:00:01}
\DefMacro{seesaw-old_D3_R4_S3_crns}{0}
\DefMacro{seesaw-old_D3_R4_S3_time}{00:00:00}
\DefMacro{seesaw-old_D1_R4_S8_crns}{0}
\DefMacro{seesaw-old_D1_R4_S8_time}{00:00:01}
\DefMacro{seesaw-old_D1_R4_S4_crns}{0}
\DefMacro{seesaw-old_D1_R4_S4_time}{00:00:01}
\DefMacro{seesaw-old_D2_R6_S4_crns}{0}
\DefMacro{seesaw-old_D2_R6_S4_time}{00:00:01}
\DefMacro{seesaw-old_D3_R6_S2_crns}{0}
\DefMacro{seesaw-old_D3_R6_S2_time}{00:00:00}
\DefMacro{seesaw-old_D1_R2_S2_crns}{0}
\DefMacro{seesaw-old_D1_R2_S2_time}{00:00:02}
\DefMacro{seesaw-old_D3_R6_S5_crns}{0}
\DefMacro{seesaw-old_D3_R6_S5_time}{00:00:09}
\DefMacro{seesaw-old_D1_R4_S6_crns}{0}
\DefMacro{seesaw-old_D1_R4_S6_time}{00:00:02}
\DefMacro{seesaw-old_D1_R2_S6_crns}{0}
\DefMacro{seesaw-old_D1_R2_S6_time}{00:00:01}
\DefMacro{seesaw-old_D3_R6_S4_crns}{0}
\DefMacro{seesaw-old_D3_R6_S4_time}{00:00:01}
\DefMacro{seesaw-old_D2_R4_S6_crns}{2}
\DefMacro{seesaw-old_D2_R4_S6_time}{00:03:30}
\DefMacro{seesaw-old_D3_R6_S7_crns}{ERROR}
\DefMacro{seesaw-old_D3_R6_S7_time}{ERROR}
\DefMacro{seesaw-old_D2_R6_S2_crns}{0}
\DefMacro{seesaw-old_D2_R6_S2_time}{00:00:00}
\DefMacro{seesaw-old_D3_R6_S3_crns}{0}
\DefMacro{seesaw-old_D3_R6_S3_time}{00:00:00}
\DefMacro{seesaw-old_D2_R4_S8_crns}{0}
\DefMacro{seesaw-old_D2_R4_S8_time}{00:00:25}
\DefMacro{seesaw-old_D2_R6_S6_crns}{0}
\DefMacro{seesaw-old_D2_R6_S6_time}{00:00:48}
\DefMacro{seesaw-old_D2_R4_S7_crns}{0}
\DefMacro{seesaw-old_D2_R4_S7_time}{00:00:16}
\DefMacro{seesaw-old_D2_R4_S3_crns}{0}
\DefMacro{seesaw-old_D2_R4_S3_time}{00:00:00}
\DefMacro{seesaw-old_D1_R6_S7_crns}{0}
\DefMacro{seesaw-old_D1_R6_S7_time}{00:00:02}
\DefMacro{seesaw-old_D3_R2_S5_crns}{0}
\DefMacro{seesaw-old_D3_R2_S5_time}{00:00:01}
\DefMacro{seesaw-old_D1_R2_S5_crns}{0}
\DefMacro{seesaw-old_D1_R2_S5_time}{00:00:01}
\DefMacro{seesaw-old_D1_R4_S3_crns}{0}
\DefMacro{seesaw-old_D1_R4_S3_time}{00:00:01}
\DefMacro{seesaw-old_D3_R4_S2_crns}{0}
\DefMacro{seesaw-old_D3_R4_S2_time}{00:00:00}
\DefMacro{seesaw-old_D3_R6_S8_crns}{ERROR}
\DefMacro{seesaw-old_D3_R6_S8_time}{ERROR}
\DefMacro{seesaw-old_D3_R4_S7_crns}{2}
\DefMacro{seesaw-old_D3_R4_S7_time}{02:26:07}
\DefMacro{seesaw-old_D1_R4_S7_crns}{0}
\DefMacro{seesaw-old_D1_R4_S7_time}{00:00:02}
\DefMacro{seesaw-old_D2_R2_S5_crns}{0}
\DefMacro{seesaw-old_D2_R2_S5_time}{00:00:01}
\DefMacro{seesaw-old_D2_R2_S8_crns}{0}
\DefMacro{seesaw-old_D2_R2_S8_time}{00:00:01}
\DefMacro{seesaw-old_D2_R6_S5_crns}{0}
\DefMacro{seesaw-old_D2_R6_S5_time}{00:00:12}
\DefMacro{seesaw-old_D2_R4_S4_crns}{0}
\DefMacro{seesaw-old_D2_R4_S4_time}{00:00:01}
\DefMacro{seesaw-old_D3_R2_S2_crns}{0}
\DefMacro{seesaw-old_D3_R2_S2_time}{00:00:00}
\DefMacro{seesaw-old_D1_R6_S4_crns}{0}
\DefMacro{seesaw-old_D1_R6_S4_time}{00:00:01}
\DefMacro{seesaw-old_D1_R2_S3_crns}{1}
\DefMacro{seesaw-old_D1_R2_S3_time}{00:00:04}
\DefMacro{seesaw-old_D3_R2_S6_crns}{0}
\DefMacro{seesaw-old_D3_R2_S6_time}{00:00:01}
\DefMacro{seesaw-old_D3_R2_S3_crns}{0}
\DefMacro{seesaw-old_D3_R2_S3_time}{00:00:00}
\DefMacro{seesaw-old_D1_R6_S3_crns}{0}
\DefMacro{seesaw-old_D1_R6_S3_time}{00:00:00}
\DefMacro{seesaw-old_D2_R6_S8_crns}{0}
\DefMacro{seesaw-old_D2_R6_S8_time}{00:07:22}
\DefMacro{seesaw-old_D1_R4_S5_crns}{0}
\DefMacro{seesaw-old_D1_R4_S5_time}{00:00:01}
\DefMacro{seesaw-old_D3_R4_S5_crns}{0}
\DefMacro{seesaw-old_D3_R4_S5_time}{00:00:04}
\DefMacro{seesaw-old_D3_R4_S4_crns}{0}
\DefMacro{seesaw-old_D3_R4_S4_time}{00:00:01}
\DefMacro{seesaw-old_D3_R2_S4_crns}{1}
\DefMacro{seesaw-old_D3_R2_S4_time}{00:00:02}
\DefMacro{seesaw-old_D1_R4_S2_crns}{0}
\DefMacro{seesaw-old_D1_R4_S2_time}{00:00:00}
\DefMacro{seesaw-old_D2_R6_S3_crns}{0}
\DefMacro{seesaw-old_D2_R6_S3_time}{00:00:01}
\DefMacro{seesaw-old_D2_R2_S4_crns}{1}
\DefMacro{seesaw-old_D2_R2_S4_time}{00:00:03}
\DefMacro{seesaw-old_D1_R2_S4_crns}{0}
\DefMacro{seesaw-old_D1_R2_S4_time}{00:00:02}
\DefMacro{seesaw-old_D2_R4_S5_crns}{2}
\DefMacro{seesaw-old_D2_R4_S5_time}{00:01:02}
 
\def\figname{Fig.}
 
\newlength{\textfloatsepsave}
\title{
  CRNs Exposed: A Method for the Systematic Exploration of Chemical Reaction Networks
}

\titlerunning{CRNs Exposed} %

\author{Marko Vasic}{The University of Texas at Austin, USA}{vasic@utexas.edu}{}{}%

\author{David Soloveichik}{The University of Texas at Austin, USA}{david.soloveichik@utexas.edu}{}{}

\author{Sarfraz Khurshid}{The University of Texas at Austin, USA}{khurshid@utexas.edu}{}{}

\authorrunning{M. Vasic and D. Soloveichik and S. Khurshid} %

\Copyright{Marko Vasic and David Soloveichik and Sarfraz Khurshid} %



\category{} %

\relatedversion{} %

\supplement{}%

\acknowledgements{This work was supported in part by NSF grants CCF-1901025 to DS and CCF-1718903 to SK}%

\nolinenumbers

\EventEditors{Cody Geary and Matthew J. Patitz}
\EventNoEds{2}
\EventLongTitle{26th International Conference on DNA Computing and Molecular Programming (DNA 26)}
\EventShortTitle{DNA 26}
\EventAcronym{DNA}
\EventYear{2020}
\EventDate{September 13--18, 2020}
\EventLocation{Oxford, UK (Virtual Conference)}
\EventLogo{}
\SeriesVolume{174}
\ArticleNo{4}

\begin{document}

\maketitle

\begin{abstract}
Formal methods have enabled breakthroughs in many fields, such as in hardware verification, machine learning and biological systems.
The key object of interest in systems biology, synthetic biology, and molecular programming is \emph{chemical reaction networks} (CRNs) which formalizes \emph{coupled chemical reactions} in a well-mixed solution.
CRNs are pivotal for our understanding of biological regulatory and metabolic networks, as well as for programming engineered molecular behavior.  
Although it is clear that small CRNs are capable of complex dynamics and computational behavior, 
it remains difficult to explore the space of CRNs in search for desired functionality. 
We use Alloy, a tool for expressing structural constraints and behavior in software systems, 
to enumerate CRNs with declaratively specified properties. 
We show how this framework can enumerate CRNs with a variety of structural constraints 
including biologically motivated catalytic networks and metabolic networks, 
and seesaw networks motivated by DNA nanotechnology. 
We also use the framework to explore analog function computation in rate-independent CRNs. 
By computing the desired output value with stoichiometry rather than with reaction rates 
(in the sense that $X \to Y+Y$ computes multiplication by $2$), 
such CRNs are completely robust to the choice of reaction rates or rate law.
We find the smallest CRNs computing the \emph{max}, \emph{minmax}, \emph{abs} and \emph{ReLU} (rectified linear unit) functions in a natural subclass of rate-independent CRNs where rate-independence follows from structural network properties.
\end{abstract}

\Section{Introduction}

Formal methods have enabled breakthroughs in many fields, e.g., in hardware verification~\cite{modelChecking}, machine learning~\cite{GehrETAL18AI2,huang2017safety}, and biological systems~\cite{bernot2004application,giacobbe2015model,heath2008probabilistic,lakin2012design,WangETALProbabilistic}.
In this paper we apply formal methods to \emph{Chemical Reaction Networks} (CRNs), which have been objects of intense study in systems and synthetic biology.
CRNs are widely used in modeling biological regulatory networks,
and essentially identical models are also widely used in ecology~\cite{volterra1927variazioni}, distributed computing~\cite{angluin2007computational}, and other fields.
More recently, CRNs have been directly used as a programming language for engineering molecules obeying prescribed interaction rules via DNA strand displacement cascades~\cite{SoloveichikETAL10DNAUniversalSubstrate,cardelli2011strand,ChenETAL13ProgrammableChemicalControllersFromDNA,srinivas2017enzyme,shah2020using}.

It is clear that small CRNs can exhibit very complex behavior. 
Dynamical systems, e.g., oscillatory, chaotic, and bistable systems, typically contain only a few reactions.
Small CRNs also exhibit interesting computational behavior.
For example, the approximate majority population protocol studied in distributed computing~\cite{angluin2008simple} was later identified with a variety of biological networks~\cite{cardelli2014morphisms}. 
Can we systematically explore the power of small reaction networks?

We present a method that exhaustively enumerates small CRNs in different classes that are relevant for biology and for synthetic engineering systems.
The enumeration is performed using Alloy, a powerful tool for modeling structural constraints and behavior in software systems using first-order logic with transitive
closure~\cite{Jackson02Alloy}.
The Alloy tool performs \emph{scope-bounded} analysis~\cite{JacksonETAL00ALCOA}.  Given an Alloy model and a
\emph{scope}, i.e., a bound on the universe of discourse, the analyzer
translates the Alloy model to a propositional satisfiability (SAT)
formula and invokes an off-the-shelf SAT solver~\cite{EenSorensson03MiniSAT} to analyze the model.
Alloy is used in a wide range of areas in
software engineering, including
software
design~\cite{JacksonFeketeTACS01,FriasETALICSE05},
analysis~\cite{JacksonVaziriISSTA00,DennisETAL06,GaleottiETALTSE13,KhurshidETAL02AnalyzableAnnotationLanguage},
testing~\cite{MarinovKhurshid01TestEra}, and
security~\cite{KangETALFSE16}.
We show how Alloy can be used to conveniently model interesting classes of CRNs for biology and bioengineering, and we use the Alloy analyzer to search for CRNs with specific desired functionality.

As examples of the method we first focus on a number of classes: elementary, catalytic, metabolic.
We say \emph{elementary reactions} are CRNs with at most two reactants and products.
(We allow reactions to be irreversible;
reversible reactions are represented by two irreversible reactions.)
\emph{Catalytic networks} are those elementary CRNs in which the reactants and products are not disjoint; i.e., the reaction is catalyzed by some species that is not consumed in the reaction.
Catalytic networks (e.g., transcriptional, phosphorylation, etc.) regulate many aspects of the cell's behavior~\cite{ptacek2005global,lee2002transcriptional}.
In general protein-protein interactions, proteins can catalytically modify other proteins, which in turn can be catalysts in other interactions.
An important subclass of catalytic networks are \emph{metabolic networks}, where the enzymes are proteins while the substrates are small molecules;  these catalytic CRNs are ``bipartite'' in the sense that a species is either always a catalyst or never a catalyst.
\emph{Autocatalytic networks} are another interesting subclass of catalytic networks in which the (auto)catalyst generates another copy of itself.
Autocatalysis is useful for exponential amplification and oscillation.

We then turn our attention to classes of CRNs especially relevant for synthetic reaction networks, showing how abstract molecular structure can be modeled in Alloy. 
In particular, we focus on DNA strand displacement cascades, which have proved to be a uniquely programmable technology for cell-free DNA-only systems~\cite{zhang2011dynamic}.
Strand displacement interactions correspond to reactions between two types of molecules: ``gates'' and ``strands'',
where the reacting strand displaces the strand previously sequestered in the gate complex.
A simple, yet very scalable, class of strand displacement circuits uses a simple motif called 
\emph{seesaw} gates~\cite{qian2011simple,qian2011scaling,cherry2018scaling}
 that makes use of a reversible
strand displacement reaction.
 We designed an Alloy model to enumerate such strand displacement reactions, showing that abstract molecular structure can be  incorporated into the Alloy modeling formalism. 

In the second part of the paper, we use our enumeration framework to search for specific desired functionality in a class of CRNs.
In particular, we focus on the class of rate-independent CRNs~\cite{chen2014rate}. 
Consider the reaction $X \to Y+Y$, and think of the concentrations of species $X$ and $Y$ as input and output respectively.
This reaction computes the function of ``multiplication by $2$'' since in the limit of time going to infinity it produces two units of $Y$ for every unit of $X$ initially present. 
Similarly the reaction $X_1 + X_2 \to Y$ computes the ``minimum'' function since the amount of $Y$ eventually produced will be the minimum of the initial amounts of $X_1$ and $X_2$.
Note that such computation makes no assumption on the rate law, such as whether the reaction obeys mass-action kinetics\footnote{``Mass-action'' kinetics refers to the best-studied case where the reaction rate is proportional to the product of the concentration of the reactants.}
or not, allowing the computation to be correct in a wide variety of chemical contexts.
(We use the continuous CRN model where concentrations are real-valued quantities.)

A natural subclass of CRNs whose structure enforces rate independence are those that satisfy two constraints: feed-forward, and non-competitive.\footnote{Feed-forward and non-competitive conditions are sufficient for rate-independence, but are not necessary. However, most known examples of rate independent computation satisfy these conditions.} 
Intuitively, the first condition ensures that the CRN converges to a static equilibrium where no reaction can occur.
The second condition ensures that no matter what the rates are, the system converges to the \emph{same} static equilibrium.
More precisely, we define feed-forward as follows:
there exists a total ordering on the reactions such that no reaction consumes\footnote{\label{fn1}We say a reaction \emph{produces} (resp.~\emph{consumes}) a species $S$ if there is net stoichiometric gain (resp.~loss) of $S$. Thus a catalyst in a reaction is neither consumed nor produced.} a species produced by a reaction later in the ordering.
We define non-competitive as follows: if a species is consumed in a reaction then it cannot appear as a reactant somewhere else. 
Such constraints on the structure of the network can be easily encoded in the Alloy specification.
We also require each reaction to consume at least one species (boundedness condition).
We show in Appendix~\ref{app:rateindependence} that these conditions ensure that the CRN is rate-independent.

\begin{figure}[!t]
  \begin{align}
    \ce{$A$ &->[] $Z_1$ + $Y$} \nonumber \\
    \ce{$B$ &->[] $Z_2$ + $Y$} \nonumber \\
    \ce{$Z_1$ + $Z_2$ &->[] $K$} \nonumber \\
    \ce{$Y$ + $K$ &->[] $\emptyset$} \nonumber
  \end{align}
\vspace{-10pt}
\Caption{CRN computing Max. We think of the initial amount of $A$ and $B$ as inputs, and the converging amount of $Y$ as the output. 
The amount of $Y$ eventually produced in reactions $1$ and $2$ is the sum of the initial amounts of $A$ and $B$. The amount of $K$ eventually produced in reaction $3$ is the minimum of the initial amounts of $A$ and $B$. Reaction $4$ subtracts the minimum from the sum, yielding the maximum.
(The $4$th reaction generates waste species, which are not named.)
}
\label{fig:max}
\vspace{-5pt}
\end{figure}

Focusing on the class of feed-forward, non-competitive CRNs, we search for the smallest reaction networks implementing \emph{max}, \emph{minmax}, \emph{abs}, and \emph{ReLU} (rectified linear unit) functions.
As an example of the kind of computation we achieve, consider the \emph{max} computing CRN shown in \figname~\ref{fig:max}.
This CRN was previously studied~\cite{chen2014deterministic,chen2014rate}; our result shows that it is indeed the smallest.
The maximum function serves an important role in rate-independent computation since together with minimum, multiplication and division by a constant it forms a complete basis set~\cite{chen2014rate,chalk2018composable}.
The \Relu function was first introduced due to the biological motivations explaining functioning of neurons in the brain cortex~\cite{hahnloser2000digital}.
Since then, it was used with great success in the machine learning community, particularly in deep learning~\cite{lecun2015deep,glorot2011deep} for realizing artificial neural networks.
The simplicity of its implementation suggests that CRNs can naturally realize neural computation~\cite{vasic2020DMP}.
To our knowledge, the smallest implementations of \emph{abs} (absolute value), and \emph{minmax} (a two output function computing both minimum and maximum of two inputs) that we find are novel and have not been previously published.

Much ongoing work explores the computational power of CRNs.
Previous work showed the implementation of numerous complex behaviors, such as mapping polynomials to
chemical reactions~\cite{SalehiETAL17CRNsForComputingPolynomials},
programming logic gates~\cite{Magnasco97ChemicalKineticsIsTuringUniversal},
mapping discrete, control flow, algorithms~\cite{HuangETAL12CompilingControlFlowIntoBiochemicalReactions},
and a molecular programming language translating high-level specifications to chemical reactions~\cite{vasic2018crn++}.
However the complexity of these reaction systems can be infeasible, asking for novel techniques that answer what is the natural way to compute ``in reactions''.
To help answer this question we can take a different, bottom-up approach, and explore what small CRNs naturally do.
We believe that insight we get from exploring reactions will help in design of higher-level primitives that naturally map to reactions,
and will provide knowledge for more efficient design of high-level languages.
We release the source code~\cite{CRNsExposedGithub} of the tool to enable others make use of it, and extend it further.

\Section{Modeling CRNs in Alloy}
\label{sec:modeling}

This section describes our approach to modeling chemical reaction networks (CRNs) in Alloy.
(See Appendix~\ref{app:alloy} for additional background on Alloy.)
We first introduce a general model to represent the broadest class of CRNs (allowing arbitrary number of reactants and products),
and next show specializations of the model for different classes such as \emph{elementary}, \emph{catalytic}, \emph{metabolic}, \emph{autocatalytic}, and \emph{feed-forward non-competitive} reactions.
Next, we present models that encode abstract molecular structure, including \emph{strands and gates} model and a \emph{seesaw} model built on top of it.
Our approach naturally admits a hierarchical structuring of models where a model builds on and specializes another model---e.g., metabolic reactions are structurally more constrained reactions than elementary.
This allows a systematic exploration of the design space of models as this section illustrates.

\textbf{General model.}
\begin{figure}[!t]
\begin{CodeOut}
\begin{verbatim}
module crn

abstract sig Species {}
abstract sig Reaction { reactants, products: seq Species }

-- Basic semantic constraints -- for all CRNs
fact AtLeastOneReactant { -- each reaction has >=1 reactant
  all r: Reaction | some r.reactants }

fact UniqueReactions { -- each reaction is unique
  all disj r1, r2 : Reaction | ReactionsDifferent[r1, r2] }

pred ReactionsDifferent[r1, r2: Reaction] {
  SpeciesSeqDifferent[r1.reactants, r2.reactants]
  or SpeciesSeqDifferent[r1.products, r2.products] }

pred SpeciesSeqDifferent[seq1, seq2: seq Species] {
  some s : Species | #indsOf[seq1, s] != #indsOf[seq2, s] }

fact ReactantsDifferentThanProducts {
  all r: Reaction | SpeciesSeqDifferent[r.reactants, r.products] }

fact AllSpeciesUsed { -- each species is used in some reaction
  Int.(Reaction.(reactants + products)) = Species }

pred ContainsAsReactant[r: Reaction, s: Species] { s in Int.(r.reactants) }
pred ContainsAsProduct[r: Reaction, s: Species] { s in Int.(r.products) }
\end{verbatim}
\end{CodeOut}
\Caption{General Alloy model of CRNs.  ``$--$'' indicate
  start of a comment\XComment{.  Alloy comments may alternatively be written
    using ``$//$'' or as a block using ``$/* ... */$''}.\label{fig:crn}}
\end{figure}
Our general model captures CRNs consisting of reactions with arbitrarily many reactants and products.
To model this in Alloy we define a set of species, a set of reactions, two relations that characterize the reactants and products,
and logical constraints that define the basic structural requirements for well-formed CRNs.
\figname~\ref{fig:crn} specifies the general model in Alloy.
The keyword \CodeIn{module} allows naming the model, which can be imported in other models.
The keyword \CodeIn{sig} declares a basic type and introduces a set of indivisible atoms that do not have any internal structure.
The model declares two sets: a set of species (\CodeIn{Species}) and a set of reactions (\CodeIn{Reaction}).
The signature declaration of \CodeIn{Reaction} introduces two \emph{fields},
\CodeIn{reactants} and \CodeIn{products}, each of type \emph{sequence}
(\CodeIn{seq}) of \CodeIn{Species}.  Alloy models a sequence as a
binary relation from (non-negative) integer indices to atoms.  Thus,
each of these field declarations introduces a ternary relation of
type: \CodeIn{Reaction $\times$ Int $\times$ Species}.
In a case of reaction $R0: X \to Y+Y$, the value of products relation would be the set: $\{R0 \times 0 \times Y, R0 \times 1 \times Y\}$.
Note that we model reactants and products with \emph{seq} instead of \emph{set} to support repetition of a species as a reactant or product, as in the above reaction.

After defining the basic structure, we use Alloy \emph{facts} to add constraints ensuring that enumerated CRNs are well-formed.
A \CodeIn{fact} paragraph states a constraint that must always be satisfied,
i.e., every solution found (CRN enumerated) must satisfy each fact (and may satisfy additional constraints as desired).
For example, the fact \CodeIn{AtLeastOneReactant} requires that every reaction contains at least one reactant.
We use universal quantification (\CodeIn{all}) to require that the reactants in each reaction form a non-empty sequence.
The keyword \CodeIn{some} in formula ``\CodeIn{some E}'' for expression \CodeIn{E} constrains it to represent a non-empty set.
The operator `\CodeIn{.}' is relational
join; specifically, if \CodeIn{r} and \CodeIn{s} are binary relations
where the domain of \CodeIn{r} is the same as co-domain of \CodeIn{s},
\CodeIn{r.s} is relational composition, and if \CodeIn{x} is a scalar
and \CodeIn{t} is a binary relation where the type of \CodeIn{x} is
the co-domain of \CodeIn{t}, \CodeIn{x.t} is relational image of
\CodeIn{x} under \CodeIn{t}.
Thus, \CodeIn{r.reactants} represents a sequence of reactants in a reaction $r$.

We ensure that there are no two identical reactions in a CRN using the fact \CodeIn{UniqueReactions}.
For all distinct (\CodeIn{disj}) reactions we require that predicate \CodeIn{ReactionsDifferent} holds.
A \emph{predicate} (\CodeIn{pred}) paragraph is a named formula that may have parameters.
The predicate \CodeIn{ReactionsDifferent} uses logical disjunction (\CodeIn{or}) and invokes \CodeIn{SpeciesSeqDifferent} to constrain its parameters (reactions) \CodeIn{r1} and \CodeIn{r2} to be different.

The predicate \CodeIn{SpeciesSeqDifferent} is true if the two sequences of species are different.
It uses existential quantification (\CodeIn{some}).
The operator `\CodeIn{\#}' represents set cardinality.
The Alloy library function \CodeIn{indsOf} represents the set of indices where the atom argument (e.g., \CodeIn{s}) appears in the sequence argument (e.g., \CodeIn{seq1}).
Intuitively, this predicate compares the number of appearances of species in two sequences, and returns true if exists a species that appears a different number of times in the two sequences.

The fact \CodeIn{ReactantsDifferentThanProducts} requires each
reaction to have non-identical reactants and products.  Finally, the
fact \CodeIn{AllSpeciesUsed} states that all species must be a part of
some reaction.
\CodeIn{Int} represents the set of integers.

The predicate \CodeIn{ContainsAsReactant} is true if a given reaction contains a given species as a reactant. Similar holds for \CodeIn{ContainsAsProduct} and reaction products.

\textbf{Illustrating the General Model.}
To illustrate using the Alloy analyzer, consider generating an
instance of the constraints modeled.  
The following \CodeIn{Generate} command instructs the analyzer to create
an instance with respect to a universe that contains exactly
2~reactions and 2~species, and 2-bit integers, and conforms to all the
facts in the model:

\begin{CodeOut}
\begin{verbatim}
Generate: run {} for exactly 2 Reaction, exactly 2 Species, 2 int
\end{verbatim}
\end{CodeOut}
Executing the command \CodeIn{Generate} and enumerating the first three instances creates the following CRNs where $S0$ and $S1$ are species,
and $\emptyset$ are waste species
\footnote{Alloy shows each instance as a valuation to the sets and relations declared in the model, and also supports visualizing the instances as graphs. We write the reactions here using their natural representation for clarity.}:

\begin{tabular}{ccc}
  \begin{minipage}{0.33\textwidth} %
  \begin{align}
    \ce{$S_1$ &->[] $S_0$} \nonumber \\
    \ce{$S_0$ &->[] $S_1$} \nonumber
  \end{align}
  \end{minipage}
  &
  \begin{minipage}{0.33\textwidth} %
  \begin{align}
    \ce{$S_1$ &->[] $\emptyset$} \nonumber \\ %
    \ce{$S_1$ &->[] $S_0$} \nonumber
  \end{align}
  \end{minipage}
  &
  \begin{minipage}{0.33\textwidth} %
    \begin{align}
      \ce{$S_1$ &->[] $\emptyset$} \nonumber \\ %
      \ce{$S_0$ &->[] $S_1$} \nonumber
    \end{align}
  \end{minipage}\\
  (a) \hspace{50pt} & (b) \hspace{50pt} & (c) \hspace{50pt} \\
\end{tabular}

While quite small, these three instances exhibit interesting properties,
CRN in (a) models a reversible reaction $S1 \longleftrightarrow S0$;
CRN in (b) is rate-dependent, where amount of $S1$ in a limit of time going to infinity is $0$,
but amount of $S0$ is dependent on reaction rates;
and CRN in (c) is rate-independent, where concentrations of both $S0$ and $S1$ converge to $0$.

\textbf{Elementary reactions.}
\begin{figure}[!t]
\begin{CodeOut}
\begin{verbatim}
module elementary
open crn
pred Elementary() { MaxReactantsNum[2] and MaxProductsNum[2] }
pred MaxReactantsNum[num: Int] { all r: Reaction | lte[#r.reactants, num] }
pred MaxProductsNum[num: Int] { all r: Reaction | lte[#r.products, num] }
\end{verbatim}
\end{CodeOut}
\Caption{Elementary reactions.\label{fig:elementary}}
\end{figure}
Elementary reactions have at most $2$ reactants and at most $2$ products.
Elementary reactions are arguably the ones commonly occurring in nature,
as it is unlikely that $3$ (or more) molecules react or split at the same exact time.
Also, reactions with more than $2$ reactants can be represented with elementary reactions;
e.g. reaction $A+B+C \to D$ can be constructed with two elementary reactions: $A+B \to T$ and $T+C \to D$.
(Similarly for products.)

\figname~\ref{fig:elementary} shows the Alloy model of \emph{elementary} reactions,
which specializes (restricts) the general CRN model \CodeIn{crn}.
The Alloy model \CodeIn{elementary} imports (\CodeIn{open}) the \CodeIn{crn} model and defines the predicate \CodeIn{Elementary},
which uses the conjunction (\CodeIn{and}) of two helper predicates \CodeIn{MaxReactantsNum} and \CodeIn{MaxProductsNum} to characterize elementary reactions.
The predicate \CodeIn{lte} is a standard Alloy utility predicate and represents the $\le$ comparison.

\textbf{Catalytic reactions.}
\begin{figure}[!t]
\begin{CodeOut}
\begin{verbatim}
module catalytic
open elementary
pred Catalytic[] { all r: Reaction | CatalyticReaction[r] }
pred CatalyticReaction[r: Reaction] { some elems[r.reactants] & elems[r.products] }
run { Catalytic and Elementary } for 2
\end{verbatim}
\end{CodeOut}
\caption{Catalytic reactions.\label{fig:catalytic}}
\end{figure}
Next, we model catalytic reactions (\figname~\ref{fig:catalytic}).  The
predicate \CodeIn{Catalytic} uses the helper predicate
\CodeIn{CatalyticReaction} to require each reaction to be catalytic,
i.e., have some species that is both a reactant and a product in that
reaction.  The Alloy utility function \CodeIn{elems} represents the
set of elements in its argument sequence; the operator `\CodeIn{\&}'
represents set intersection.
The \CodeIn{run} command instructs the
analyzer to create an instance that is both a catalytic and an
elementary reaction within a scope of 2, i.e., at most 2 atoms in each
sig.  An example instance created by executing the command is:
\begin{align}
    \ce{$S_0 + S_1 \to S_0 + S_0$}\nonumber\\
    \ce{$S_0 + S_1 \to S_1 + S_1$}\nonumber
\end{align}
We also model autocatalytic reactions shown in Appendix~\ref{app:autocatalytic}.

\textbf{Metabolic reactions.}
\begin{figure}[!t]
\begin{CodeOut}
\begin{verbatim}
module metabolic
open catalytic

pred Metabolic[] {
  Catalytic[] and
  all s: Species | (some r: Reaction | IsCatalyst[s, r]) implies
                   all x: Reaction | Contains[x, s] implies IsCatalyst[s, x] }

pred IsCatalyst[s: Species, r: Reaction] { s in Int.(r.reactants) & Int.(r.products) }
pred Contains[r: Reaction, s: Species] { ContainsAsReactant[r, s] or ContainsAsProduct[r, s] }
\end{verbatim}
\end{CodeOut}
\Caption{Metabolic reactions.\label{fig:metabolic}}
\end{figure}
In metabolic networks catalysts are proteins that act upon substrates that are small molecules. 
Thus metabolic reactions are a form of catalytic reactions in which if a species appears as a catalyst in a reaction,
then it has to be a catalyst in all reactions in which the species occurs. The predicate \CodeIn{Metabolic} in \figname~\ref{fig:metabolic} specifies metabolic reactions.
\XComment{Note how we use Int.(r.reactants) in metabolic code, while we used elems[r.reactants] in elementary reactions;
  Consider switching everywhere to Int.}

\textbf{Strands and gates.}
\begin{figure}[!t]
\begin{CodeOut}
\begin{verbatim}
module strandsandgates
open crn

sig Strand, Gate extends Species {}
fact { Strand + Gate = Species } -- strands and gates partition species

pred StrandsAndGates() {
  ExactReactantsNum[2] and ExactProductsNum[2] and
  all r: Reaction {
    some Int.(r.reactants) & Strand and some Int.(r.reactants) & Gate
    some Int.(r.products) & Strand and some Int.(r.products) & Gate }}

pred ExactReactantsNum[num: Int] { all r: Reaction | eq[#r.reactants, num] }
pred ExactProductsNum[num: Int] { all r: Reaction | eq[#r.products, num] }
\end{verbatim}
\end{CodeOut}
\Caption{Strands and gates.\label{fig:strandsandgates}}
\end{figure}
We next model synthetic CRNs which use DNA strand displacement cascades for its implementation.
Strand displacement interactions correspond to reactions between two types of molecules: ``gates'' and ``strands'',
where the reacting strand displaces the strand previously sequestered in the gate complex.
We first capture the bipartite nature of the reactions:
\figname~\ref{fig:strandsandgates} declares strands and gates as disjoint subsets (\CodeIn{extends}) that partition species.
The predicate \CodeIn{StrandsAndGates} requires that each reaction has exactly 2 reactants and 2 products,
and moreover has a strand and a gate as a reactant, and a strand and a gate as a product.

\begin{figure}[!t]
\centering
\includegraphics[scale=0.5]{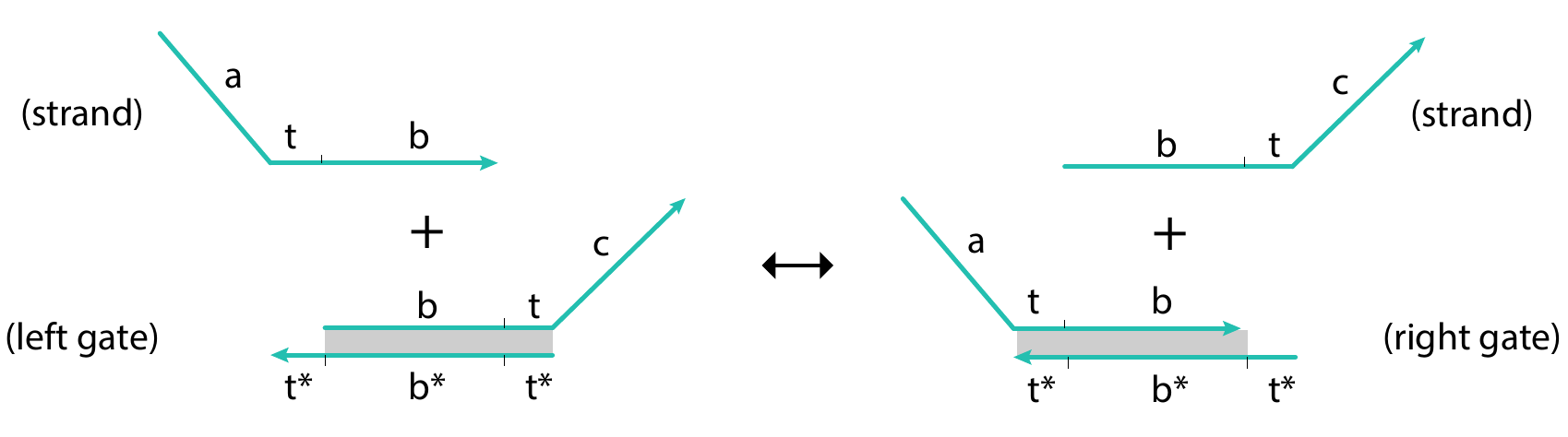}
\Caption{DNA strand displacement reaction with the seesaw gate motif.
There are two reactants (a strand and a gate) and two products (a strand and a gate). 
A gate consists of two strands bound together. 
(For simplicity the usual helical structure of DNA is not shown.)
Labels show binding sites (domains);
a star indicates Watson-Crick complement such that domain $x$ binds $x^*$.
In order for the reaction to happen, the complementary domains must match as shown.
Such reactions can be cascaded since the strands $<a,t,b>$ and $<b, t, c>$ can react with other seesaw gates.
}
\label{fig:seesaw-reaction}
\end{figure}
\textbf{Seesaw networks.}
\begin{figure}[!t]
\begin{CodeOut}
\begin{verbatim}
open strandsandgates

sig Domain {}
sig DNASpecies in Species { left, right: lone Domain }
sig RightGate, LeftGate extends Gate {}

fact UseAll { DNASpecies = Species and DNASpecies.(left + right) = Domain }
fact UniqueSpecies {
  all s1, s2: Strand | s1.left = s2.left and s1.right = s2.right implies s1 = s2
  all s1, s2: RightGate | s1.left = s2.left and s1.right = s2.right implies s1 = s2
  all s1, s2: LeftGate | s1.left = s2.left and s1.right = s2.right implies s1 = s2 }
fact OneDomain { all s: Strand + LeftGate + RightGate | one s.left and one s.right }

pred CanReactStrandAndLeftGate[s: Strand, lg: LeftGate] {
  s in Strand and lg in LeftGate and s.right = lg.left }
pred CanReactStrandAndRightGate[s: Strand, rg: RightGate] {
  s in Strand and rg in RightGate and s.left = rg.right }
pred CanReact[r1: DNASpecies, r2: DNASpecies] {
  CanReactStrandAndLeftGate[r1, r2] or CanReactStrandAndRightGate[r1, r2] }

pred ReactStrandAndLeftGate[s: Strand, lg: LeftGate, s':Strand, rg': RightGate] {
  (s in Strand and lg in LeftGate and s' in Strand and rg' in RightGate
   and CanReactStrandAndLeftGate[s, lg] 
   and s'.left = lg.left and s'.right = lg.right and rg'.left = s.left and rg'.right = s.right) }
pred ReactStrandAndRightGate[s: Strand, rg: RightGate, s': Strand, lg': LeftGate] {
  (s in Strand and rg in RightGate and s' in Strand and lg' in LeftGate
   and CanReactStrandAndRightGate[s, rg]
   and s'.left = rg.left and s'.right = rg.right and lg'.left = s.left and lg'.right = s.right) }
pred React[r1: Species, r2: Species, p1: Species, p2: Species] {
  ReactStrandAndLeftGate[r1, r2, p1, p2] or ReactStrandAndRightGate[r1, r2, p1, p2] }

fun ReactantsSet[r: Reaction]: set Species { Int.(r.reactants) }
fun ProductsSet[r: Reaction]: set Species { Int.(r.products) }

pred Seesaw {
  StrandsAndGates[]
  all r: Reaction { -- All reactions are seesaw reactions.
    let s = 0.(r.reactants), g = 1.(r.reactants), s' = 0.(r.products), g' = 1.(r.products) {
       React[s, g, s', g'] }}
  all s1, s2: Species { -- All possible reactions exist.
    CanReact[s1, s2] implies some r: Reaction {
      (s1 + s2) = ReactantsSet[r] or (s1 + s2) = ProductsSet[r] }}
  all s1, s2: Species | all rxn1, rxn2: Reaction { -- Prevent reverse direction.
    ((s1+s2) = ReactantsSet[rxn1]) implies ((s1+s2) != ProductsSet[rxn2]) }
  all r: Reaction { some LeftGate & ReactantsSet[r] }
}

GenSeesaw: run Seesaw for exactly 1 Reaction, exactly 3 Domain, exactly 4 Species
\end{verbatim}
\end{CodeOut}
\Caption{Seesaw model.\label{fig:seesaw}}
\vspace{-5pt}
\end{figure}
A simple yet powerful subclass of DNA strand displacement reactions is the ``seesaw'' model.
Seesaw reactions have been used to create some of the largest synthetic biochemical reaction networks, including logic circuits and neural networks~\cite{qian2011scaling,cherry2018scaling}.
The molecular structure schematic for a seesaw reaction is shown in \figname~\ref{fig:seesaw-reaction}.
\figname~\ref{fig:seesaw} models seesaw reactions by specializing the model of strands and gates (\figname~\ref{fig:strandsandgates}),
capturing the abstract molecular structure in an Alloy model.
The signature \CodeIn{Domain} models the binding domains.
The signature \CodeIn{DNASpecies} is a subset (\CodeIn{in}) of species,
and \CodeIn{left} and \CodeIn{right} are binary relations that map \CodeIn{DNASpecies} to their left and right domains respectively.
The keyword \CodeIn{lone} constraints the relations to be partial functions.
The signatures \CodeIn{RightGate} and \CodeIn{LeftGate} partition gates.
The fact \CodeIn{UseAll} requires all species to be DNA species, and requires all domains to be a part of some species\XComment{ (this prevents generating instances with `hanging' domains)}.
The fact \CodeIn{UniqueSpecies} enforces that strands and gates are unique, i.e., there cannot be two or more strands (or left/right gates) with matching left and right domains.
The fact \CodeIn{OneDomain} requires strands and gates to have exactly one left and exactly one right domain.
The predicate \CodeIn{CanReactStrandAndLeftGate} is true if inputs (reactants) conform to the interaction rules of a strand and a left gate,
similar holds for the predicate \CodeIn{CanReactStrandAndRightGate} on strands and right gates.
The predicate \CodeIn{CanReact} is true if inputs (reactants) satisfy either \CodeIn{CanReactStrandAndLeftGate} or \CodeIn{CanReactStrandAndRightGate}.
The predicate \CodeIn{ReactStrandAndLeftGate} is true if inputs (reactants and products) conform to the interaction rules of a strand and a left gate,
specifically \CodeIn{s} and \CodeIn{lg} interact, i.e., the right domain of \CodeIn{s} matches the left domain of \CodeIn{lg}, and produce \CodeIn{s'} and \CodeIn{rg'} where the left and right domains of \CodeIn{s'} match those of \CodeIn{lg}, and left and right domains of \CodeIn{rg'} match those of \CodeIn{s};
likewise, \CodeIn{ReactStrandAndRightGate} specifies the interaction of a strand and a right gate.
The functions \CodeIn{ReactantsSet} and \CodeIn{ProductsSet} returns a set of reactants (products) in a reaction.
The predicate \CodeIn{Seesaw} specifies:
(a) each reaction to be a seesaw reaction by enforcing the predicate \CodeIn{React} on every reaction;
(b) that all possible reactions exist, i.e., if two species can interact based on seesaw interaction rules (predicate \CodeIn{CanReact}) than a reaction containing those species as reactants (or products) must exist;
(c) that reactions only in one direction exist (to reduce number of solutions we enforce that only one direction of reaction exist in enumerated CRNs knowing that seesaw reactions are always reversible);
(d) that reactions have a left gate as a reactant (this is to prevent multiple redundant solutions, since all reactions are reversible we can enforce that left gate is always on the left hand side).

An instance generated by Alloy running the predicate with command \CodeIn{GenSeesaw} is
$S_{ab} + LG_{bc} \to S_{bc} + RG_{ab}$,
where $S_{ab}$ and $S_{bc}$ are strands, $LG_{bc}$ left gate, $RG_{ab}$ right gate, while left and right domains $\{a,b,c\}$ are denoted in subscript.
Note that this reaction is equivalent to the one shown in \figname~\ref{fig:seesaw-reaction}.

To reduce the enumeration overhead for seesaw, we updated the \CodeIn{Reaction} signature by removing the representation of reactants and products as a sequence (sequence introduces integers as an overhead),
and adding two relations for reactants and products (as seesaw reactions are restricted to two reactants and two products). The updated \CodeIn{Reaction} signature is:
\CodeIn{abstract sig Reaction \{ r1, r2, p1, p2: Species \}}

\textbf{Feed-forward, non-competitive CRNs.}
\begin{figure}[!t]
\begin{CodeOut}
\begin{verbatim}
open elementary

one sig Graph { edges: Reaction -> Reaction }
{ all r1, r2: Reaction | r1->r2 in edges implies some s: Species |
    NetProduces[r1, s] and NetConsumes[r2, s]
  all s: Species | all r1, r2: Reaction |
    NetProduces[r1, s] and NetConsumes[r2, s] implies r1->r2 in edges }

pred DAG[] { all r: Reaction | r !in r.^(Graph.edges) }

pred NonCompetitive[] {
  all r1, r2: Reaction | all s : Species {
    (ContainsAsReactant[r1, s] and NetConsumes[r2, s]) implies r1 = r2 }}

pred NetProduces[r: Reaction, s: Species] { -- r net produces s
  lt[#indsOf[r.reactants,s], #indsOf[r.products,s]] }

pred NetConsumes[r: Reaction, s: Species] { -- r net consumes s
  gt[#indsOf[r.reactants,s], #indsOf[r.products,s]] }

pred MustConsume[] {
  all r: Reaction | some s: Species | NetConsumes[r, s] }

pred Feedforward[] { Elementary[] and DAG[] and NonCompetitive[] and MustConsume[] }
\end{verbatim}
\end{CodeOut}
\Caption{Feed-forward, non-competitive CRNs in Alloy.\label{fig:feedforward}}
\end{figure}
\figname~\ref{fig:feedforward} models feed-forward, non-competitive CRNs.
Recall, we define feed-forward as:
there exists a total ordering on the reactions such that no reaction consumes a species produced by a reaction later in the ordering.
Also, we define non-competitive as: every species is consumed by at most one reaction.

To model feed-forward constraints, one approach is to directly enforce a total ordering on the reactions with respect to the feed-forward property.
Observe that there can be multiple valid total orderings of reactions for the same feed-forward CRN,
which means that when enumerating instances for the resulting model,
multiple unique instances are created for the same CRN.
This is useful when finding all total orderings that exist for a CRN.
However, our goal is to search for CRNs exhibiting desired functionality,
and thus we aim to enumerate each CRN once, and as quickly as possible.
To tackle this problem we achieve the total ordering by creating a graph of reaction dependencies, and enforce it to be \emph{directed-acyclic}.

Our modeling of feed-forward constraints introduces a new \emph{singleton} (\CodeIn{one}) sig, termed \CodeIn{Graph},
to model a dependency relation, termed \CodeIn{edges}, between reactions.
The constraint paragraph that immediately follows the signature declaration implicitly introduces a fact that defines the edges.
Specifically, there is an edge from reaction \CodeIn{r1} to reaction \CodeIn{r2} if and only if there is some species \CodeIn{s} such that \CodeIn{r1} produces \CodeIn{s} and \CodeIn{r2} consumes \CodeIn{s}.  Total ordering is achieved by the predicate \CodeIn{DAG} that requires the graph to be \emph{directed-acyclic}.
The operator `\CodeIn{\^{}}' is transitive closure and \CodeIn{r.\^{}(Graph.edges)} represents the set of all reactions that are reachable from \CodeIn{r}.
The predicate \CodeIn{NonCompetitive} enforces that if a species is used as a reactant in a reaction then it cannot be consumed by any other reaction.
The predicate \CodeIn{MustConsume} enforces that every reaction consumes some species (boundedness condition).
The predicate \CodeIn{Feedforward} defines elementary, feed-forward, and non-competitive reactions where each reaction must consume some species.

\Section{CRN Enumeration and Search}
\label{sec:exhaustive-search}

In this section we describe our algorithm (shown in Algorithm~\ref{alg:max-search}) that performs a bounded exhaustive search enumerating all CRNs in a given class and within a given bounds respecting properties defined by an Alloy model, to find the CRN implementing desired function.

\setlength{\textfloatsep}{5pt}
\begin{algorithm}[!t]
\footnotesize
\caption{Search Algorithm}\label{alg:max-search}
\hspace*{\algorithmicindent} \textbf{Input:} Model (\textit{model}), Generation bounds (\textit{scope}), Function ($f$), Inputs ($N$). \\
\hspace*{\algorithmicindent} \textbf{Output:} CRN that computes $f$ if found; otherwise, null.
\begin{algorithmic}[1]
\Procedure{ExhaustiveSearch}{}
\ForEach {$instance \in Alloy.findAllInstances(model, scope)$}
\State $crn \gets translate(instance)$
\IfThen{$ComputesF(crn, f, N)$}{\Return $crn$}
\EndFor
\State \Return $null$
\EndProcedure
\end{algorithmic}
\end{algorithm}

Inputs to the algorithm are the Alloy model, the size of CRNs (e.g., number of reactions and species) defined by the \textit{scope}, desired target function $f$, and the number of inputs to the function $N$.
Function \textit{findAllInstances} accepts the Alloy model definition and scope, and enumerates all possible instances that satisfy the Alloy model.
\XComment{Note that we can instead write findNextInstance, and terminate when max is found, instead of invoking findAll.}
\XComment{Alloy uses a SAT solver as a back-end, which upon enumerating an instance, adds the negated formula of the instance to the set of formulas in order to find a different instance,
this is repeated until the solver returns UNSAT solution, ensuring that all possible instances are enumerated.}
Each Alloy instance is translated to CRN (step $3$).
Then, in step $4$ we invoke the Algorithm~\ref{alg:computesF} (Section~\ref{sec:checking}) to check if CRN computes $f$.
If CRN implementing given function is found then it is returned (step $4$).
If after checking all instances no satisfying CRN is found then the procedure returns \textit{null}.

\textbf{Bounded exhaustive search}.
To find the smallest CRN computing $f$ we conduct a bounded exhaustive search.
Our goal is to find a smallest (in terms of numbers of species and reactions) feed-forward, non-competitive CRN that computes $f$.
We use \emph{iterative deepening}~\cite{holzmann2004spin,godefroid1997verisoft,havelund2000model}
where we start from a small scope and iteratively increase it to a larger scope until a desired CRN is found,
where for each scope we invoke Algorithm~\ref{alg:max-search}.

\Section{CRN Analysis}
\label{sec:checking}

In this section we describe our algorithm for checking if a CRN computes a function of interest ($f$).

\textbf{Conservation Equations.}
We first construct a set of conservation equations for the CRN which describe concentrations of species in terms of their initial concentrations and reaction fluxes.
A reaction flux is equal to the total ``flow of material'' through the reaction.
We associate a flux variable to the each reaction, where $flux_i$ represents the flux of the reaction $i$.
Then the concentration of a species $S$ can be expressed in terms of its initial concentration $S_0$ and reaction fluxes:
\newcommand{\flux}{\text{\emph{flux}}}
\begin{small}
\begin{equation}
  s = s_0 + \sum_{i=1}^N netGain(rxn_i, S) \cdot \flux_i \label{eq:fluxes}
\end{equation}
\end{small}%
where $netGain(rxn_i, S)$ is the net stoichiometric gain of species $S$ in the reaction $i$ (negative in the case of loss),
and $N$ is the number of reactions in the CRN.
For example, the CRN from \figname~\ref{fig:max} generates the equations shown in~\ref{eq:max-linear}.
The variables on the left side of equations represent concentrations of species,
variables with suffixes $0$ represent initial concentrations of species
(e.g., $z_{10}$ is initial concentration of species $Z_1$),
and finally $flux_i$ variables represent fluxes of reactions.
\begin{small}
\begin{equation} \label{eq:max-linear}
\begin{tabular}{ll}
  $a = a_0 - flux_1$ & $b = b_0 - flux_2$ \\
  $z_1 = z_{10} + flux_1 - flux_3$ & $z_2 = z_{20} + flux_2 - flux_3$ \\
  $k = k_0 + flux_3 - flux_4$ & $y = y_0 + flux_1 + flux_2 - flux_4$
\end{tabular}
\end{equation}
\end{small}

\textbf{Equilibrium Condition.}
We next use the above conservation equations to find equilibria.
Since we focus on rate-independent computation, we search for static equilibria only (none of the reactions is occurring).\footnote{In chemical kinetics, \emph{static} equilibrium refers to an equilibrium where none of the reactions occur. In contrast, in  \emph{dynamic} equilibria, concentrations don't change over time because the effects of the different reactions cancel out. Note that dynamic equilibria are not rate-independent since changing a reaction rate affects the equilibrium concentrations of the species involved in that reaction.} 
A static equilibrium corresponds to every reaction having at least one reactant in zero concentration. 
Thus, we create multiple systems of equations from the conservation equations, where each system corresponds to setting concentrations of a set of species to zero, where the set contains a reactant from each reaction.
The solution of each such constructed system of equations represents concentrations of species at an equilibrium.
Different equilibria will be reached from different initial conditions.

As an example, consider again the CRN shown in \figname~\ref{fig:max}. 
All combinations of species containing a reactant from each reaction are:
{$(A, B, Z_1, Y)$, $(A, B, Z_2, Y)$, $(A, B, Z_1, K)$, $(A, B, Z_2, K)$}.
For each combination we set its species concentrations to zero and solve the system~\ref{eq:max-linear}.
This results in $4$ solutions shown in~\ref{eq:max-linear-solved} (we do not show solutions for flux variables due to the space limits).
\begin{small}
\begin{equation} \label{eq:max-linear-solved}
  \begin{tabular}{l|l|l|l|l|l}
    \boldmath${a}$ & \boldmath$b$ & \boldmath$k$ & \boldmath$y$ & \boldmath$z_1$ & \boldmath$z_2$ \\
    \hline
    $0$ & $0$ & $-b_0 + k_0 - y_0 + z_{10}$ & $0$ & $0$ & $-a_0 + b_0 - z_{10} + z_{20}$ \\ %
    $0$ & $0$ & $-a_0 + k_0 - y_0 + z_{20}$ & $0$ & $a_0 - b_0 + z_{10} - z_{20}$ & $0$ \\ %
    $0$ & $0$ & $0$ & $b_0 - k_0 + y_0 - z_{10}$ & $0$ & $-a_0 + b_0 - z_{10} + z_{20}$ \\ %
    $0$ & $0$ & $0$ & $a_0 - k_0 + y_0 - z_{20}$ & $a_0 - b_0 + z_{10} - z_{20}$ & $0$ \\ %
  \end{tabular}
\end{equation}
\end{small}
Although there are 4 solutions, for any particular initial concentrations of the species only one of the solutions is non-negative (concentrations of species must be non-negative), and thus feasible.

\textbf{Check whether CRN computes $f$.}
We then check if the equilibrium solutions are equivalent to $f$.
In general, we do not know which species correspond to the input and which to the output,
and thus we need to check for all possible combinations of the input and the output species.
First, we construct all input $n$-tuples without repeating elements from a set of species (where $n$ is the number of the inputs to $f$)\footnote{An input tuple ($a$,$b$) will be separately considered from ($b$,$a$). However, if the sought function is known to be commutative than the order of species can be ignored.}.
Second, for all species that are not in the input tuple we set initial concentrations to zero.
Third, for the output species we try any of the remaining species.
Fourth, for a given set of input and output species, we construct a piecewise function, where each solution is valid if concentrations of species are non-negative.
Finally, we use Mathematica's constraint solving procedure \emph{FindInstance} to check if the constructed piecewise function differs from function $f$.

To illustrate on our example, consider setting input species to $A$ and $B$, and output to $Y$.
The system of equations~\ref{eq:max-linear-solved} reduces to the system~\ref{eq:max-linear-solved-reduced}.
\begin{small}
\begin{equation} \label{eq:max-linear-solved-reduced}
  \begin{tabular}{l|l|l|l|l|l}
    \boldmath${a}$ & \boldmath$b$ & \boldmath$k$ & \boldmath$y$ & \boldmath$z_1$ & \boldmath$z_2$ \\
    \hline
    $0$ & $0$ & $-b_0$ & $0$ & $0$ & $-a_0 + b_0$ \\ %
    $0$ & $0$ & $-a_0$ & $0$ & $a_0 - b_0$ & $0$ \\ %
    $0$ & $0$ & $0$ & $b_0$ & $0$ & $-a_0 + b_0$ \\ %
    $0$ & $0$ & $0$ & $a_0$ & $a_0 - b_0$ & $0$ \\ %
  \end{tabular}
\end{equation}
\end{small}%
The first two solutions are infeasible since they result in species $k$ having negative concentration, $-b_0$ and $-a_0$.
More precisely they are feasible only in the trivial case where $a_0=0 \land b_0=0$.
The third solution is feasible when $b_0 \ge a_0$, in which case $y = b_0$;
while fourth solution is feasible when $a_0 \ge b_0$, in which case $y = a_0$.
Thus, we can construct the piecewise function unifying multiple equilibrium solutions into a single function:
\begin{small}
\[ y = \begin{cases} 
  b_0 & b_0\geq a_0 \\
  a_0 & a_0\geq b_0 \\
\end{cases}
\]
\end{small}%
Next, once we constructed the equilibrium piecewise function ($y(a_0,b_0)$) we invoke the Mathematica's constraint solving procedure FindInstance to find an assignment of inputs ($a_0,b_0$) for which $y$ differs from $f$,
with additional condition that initial concentrations are non-negative ($a_0 \geq 0 \land b_0 \geq 0$).
If no counterexample is found, then the CRN computes $f$ and we have finished our search.
On the other hand, if a counterexample is found, then we repeat the procedure for the next combination of input and output species.
When the list of input and output combinations is exhausted we can conclude that the CRN does not compute $f$.

\setlength{\textfloatsep}{0pt}
\begin{algorithm}[!t]
\footnotesize
\caption{ComputesF}\label{alg:computesF}
\hspace*{\algorithmicindent} \textbf{Input:} CRN \textit{crn}, Function $f$, Number of inputs $N$. \\
\hspace*{\algorithmicindent} \textbf{Output:} \textit{True} if \textit{crn} computes $f$; \textit{false} otherwise.
\begin{algorithmic}[1]
  \Procedure{ComputesF}{}
  \State $conservationEquations \gets constructConservationEquations(crn)$
  \State $equilibriumSolutions \gets \emptyset$
  \ForEach {$speciesSet \in getAllReactantCombinations(crn)$}
  \State $equilibriumEquations \gets setConcToZero(conservationEquations, speciesSet)$
  \State $solution \gets solve(equilibriumEquations)$
  \State $equilibriumSolutions.add(solution)$
  \EndFor
  \ForEach {$\{x_1,x_2,...,x_N,y\} \in getInputOutputSpecies(crn, N)$}
  \State $nonInputSpecies \gets getOtherSpecies(crn, \{x_1,x_2,...,x_n\})$
  \State $newSols \gets setInitialConcToZero(equilibriumSolutions, nonInputSpecies)$
  \State $pwF \gets constructPiecewise(newSols, y)$
  \State $counterExample \gets FindInstance(pwF \neq f(x_1,x_2,...,x_N))$
  \IfThen{$counterExample = null$}{\Return $true$}
  \EndFor
  \State \Return $false$
  \EndProcedure
\end{algorithmic}
\end{algorithm}

\textbf{Algorithm.} We implement this functionality in Mathematica by defining \textit{ComputesF} function described in Algorithm~\ref{alg:computesF}.
In step $2$, conservation equations are constructed,
while in step $3$ we initialize a set of equilibrium solutions $equilibriumSolutions$ to an empty set.
In steps $4$--$8$, we iterate over all existing sets of species containing at least one reactant from each reaction.
Specifically, function \textit{getAllReactantCombinations} computes Cartesian product over sets of reactants from different reactions;
and removes elements with the same sets of species.
In step $5$ we update the conservation equations by setting $speciesSet$ concentrations to zero, and save the linear system in $equilibriumEquations$.
In steps $6$--$7$ we solve the system of linear equations and add it to the list of equilibrium solutions
(note that since we are focused on feed-forward non-competitive reactions, a unique solution will always exist).
Next, we iterate over all combinations of input and output species $\{x_1,x_2,...,x_N,y\}$,
where $x_1$, $x_2$, ..., $x_N$ represent input species, and $y$ output species.
In step $10$ we get all the species that are not in the input species set.
In step $11$ we modify the equilibrium solutions by setting initial concentrations of $nonInputSpecies$ to zero, and we save the result in $newSols$.
In step $12$ we construct a piecewise function $pwF$ out of $newSols$.
Finally, in step $13$ we invoke the \textit{FindInstance} method to find input values for which $pwF$ is different then \emph{$f$}.
If such solution is not found then $counterExample$ is $null$, and constructed $pwF$ is implementing \emph{$f$}; in which case procedure returns \textit{true}.
If counterexample is found then the same steps are repeated for different set of input and output species.
Finally, if all combinations are exhausted procedure returns \textit{false}.

\setlength{\textfloatsep}{\textfloatsepsave}

\Section{New Results}
\label{sec:results}

In this section we present new discoveries made using the proposed techniques.
We focus on the class of feed-forward, non-competitive CRNs since they are always rate-independent. 

\textbf{Smallest \emph{max} CRN.}
We perform bounded exhaustive search for $1$ to $4$ reactions, and $1$--$6$ species,
starting with smaller number of species and reactions, and iteratively increasing the scope until the \emph{max} is found.
Table~\ref{tb:enumcount} shows the number of enumerated CRNs and Alloy enumeration time for different scope sizes.
We perform (not perfect) isomorphic breaking in Alloy by requiring lexicographic ordering on reactions among other things (details of symmetry breaking are shown in Appendix~\ref{app:symmetry}).
Note that while we perform some isomorphic breaking\footnote{Alloy can generate \emph{isomorphic} instances, i.e., two instances that are distinct but there exists a permutation on atoms, which maps one instance to the other}, not all isomorphic cases are pruned,
and thus number of non-isomorphic instances may be less then numbers reported in Table~\ref{tb:enumcount}.
In spite of this, our approach is still exhaustive, meaning that all possible CRNs will be enumerated, but some may be enumerated multiple times.
The first occurrence of \emph{max} is found in the scope of $4$ reactions and $6$ species, and it was the $124,118^{th}$ instance Alloy enumerated in that scope.
The CRN discovered is equivalent to the one shown in \figname~\ref{fig:max}, modulo reaction and species ordering.

\setlength{\textfloatsep}{\textfloatsepsave}

\BeginTable
  \begin{center}
\begin{tabular}{c|rr|rr|rr|rr}
  \toprule
  & \TNC{2}{1 \emph{Reaction}}  & \TNC{2}{2 \emph{Reactions}} & \TNC{2}{3 \emph{Reactions}} & \TNC{2}{4 \emph{Reactions}} \\
  \midrule
  1 \emph{Species} & 3    &  00:00:00   & 0     &  00:00:00           & 0       &  00:00:00         & 0          &  00:00:00 \\
  2 \emph{Species} & 10   &  00:00:00   & 22    &  00:00:00           & 0       &  00:00:00         & 0          &  00:00:00 \\
  3 \emph{Species} & 6    &  00:00:00   & 199   &  00:00:00           & 287     &  00:00:00         & 0          &  00:00:00 \\
  4 \emph{Species} & 1    &  00:00:00   & 391   &  00:00:00           & 4,666   &  00:00:05         & 5,643      &  00:00:07 \\
  5 \emph{Species} & 0    &  00:00:00   & 291   &  00:00:00           & 17,509  &  00:00:19         & 140,064    &  00:03:57 \\
  6 \emph{Species} & 0    &  00:00:00   & 100   &  00:00:00           & 27,257  &  00:00:32         & 817,742    &  00:30:35 \\
  \bottomrule
\end{tabular}
   \end{center}
  \Caption{Number of enumerated feed-forward, non-competitive CRNs\XComment{ for various scopes} and wall-clock times (hh:mm:ss) for the enumeration procedure.}
  \label{tb:enumcount}
\EndTable

\textbf{Dual-rail convention.}
Concentrations of species are always non-negative, making it impossible to represent negative values directly.
However, there is a natural way to extend computation semantics to negative values.
Instead of using a single species to represent a value, in dual-rail convention a value is represented by a difference between a two species
(e.g., the output value is equal to the concentration of species $Y^+$ minus that of $Y^-$).

An additional requirement for CRN modules is to be composable, in the sense that the output of one can be input to another.
Note, for example, that the \emph{max} system (\figname~\ref{fig:max}) is not composable because the downstream module might consume some amount of $Y$ before it is consumed in its interaction with $K$ (last reaction). 
Composability can be ensured if the output species are never consumed~\cite{chalk2018composable,severson2019composable,chugg2018output}. 
Note that consuming $Y^+$ is logically equivalent to producing $Y^-$ (and vise versa for $Y^-$), and thus we restrict dual-rail computation in this way without losing expressibility.

\begin{small}
\begin{figure}[!t]
  \begin{subfigure}[!t]{0.3\textwidth}
    \begin{align}
      \ce{$X^+$ &->[] $M$ + $Y^+$} \nonumber \\
      \ce{$M$ + $X^-$ &->[] $Y^-$} \nonumber
    \end{align}
  \end{subfigure}
  \begin{subfigure}[!t]{0.3\textwidth}
    \begin{align}
      \ce{$X^+$ &->[] $Y^+$ + $C$} \nonumber \\
      \ce{$X^-$ &->[] $Y^+$ + $E$} \nonumber \\
      \ce{$C$ + $E$ &->[] 2$Y^-$} \nonumber
    \end{align}
  \end{subfigure}
    \begin{subfigure}[!t]{0.3\textwidth}
    \begin{align}
      \ce{$X_1^+$ &->[] $M_1$ + $Y_{max}^+$} \nonumber \\
      \ce{$X_1^-$ &->[] $M_2$ + $Y_{min}^-$} \nonumber \\
      \ce{$X_2^+$ &->[] $M_2$ + $Y_{max}^+$} \nonumber \\
      \ce{$X_2^-$ &->[] $M_1$ + $Y_{min}^-$} \nonumber \\
      \ce{$M_1$ + $M_2$ &->[] $Y_{max}^-$ + $Y_{min}^+$} \nonumber      
    \end{align}
    \end{subfigure}
  \vspace{3pt}
  \Caption{Minimal \emph{\Relu} (left), \emph{abs} (middle) and \emph{minmax} (right) CRNs.
    (left) The \emph{\Relu} CRN produces $x^+(0)$ amount of $M$ and $Y^+$ by the first reaction.
    The second reaction produces $min(x^+(0),x^-(0))$ amount of $Y^-$.
    Thus, the amount of output produced is: $y=y^+-y^-=x^+(0)-min(x^+(0),x^-(0))$ which can be shown to be equal to $ReLU(x^+(0)-x^+(0))=ReLU(x)$.
    (middle) The \emph{abs} CRN produces $x^+(0)$ amount of $C$ and $E$ by the first and second reactions, respectively,
    $x^+(0)+x^-(0)$ amount of $Y^+$, and $2min(x^+(0),x^-(0))$ amount of $Y^-$.
    Thus, $y=x^+(0)+x^-(0)-2min(x^+(0),x^-(0))=abs(x^+(0),x^-(0))=abs(x)$.
  }
  \label{fig:min-crns}
\end{figure}
\end{small}

\textbf{Smallest \emph{\Relu} CRN.}
Using the above described procedure we run experiments for finding the smallest CRN computing \Relu (rectified linear unit) function.
We confirm that the CRN introduced in \cite{vasic2020DMP}, which is shown in \figname~\ref{fig:min-crns}, is indeed the smallest.
Note that CRNs were already enumerated when searching for \emph{max}, and that was no need to re-enumerate them as they were saved on disk.

Our analysis shows that the \Relu CRN is the smallest in the sense that there is no other CRN computing this function with fewer than $2$ reactions or $5$ species. 
In Appendix~\ref{app:reluminimality} we argue that our enumeration in Table~\ref{tb:enumcount} is sufficient to ensure that $5$ species are necessary no matter how many reactions are allowed.

\textbf{Smallest \emph{abs} CRN.}
We conducted a similar experiment for finding the smallest CRN computing the absolute value function, 
finding CRN shown in \figname~\ref{fig:min-crns}.

\BeginTable
  \begin{center}
\begin{tabular}{c|rr|rr|rr|rr}
  \toprule
          & \TNC{2}{2 \emph{Reactions}} & \TNC{2}{3 \emph{Reactions}} & \TNC{2}{4 \emph{Reactions}} & \TNC{2}{5 \emph{Reactions}} \\
  \midrule
  8 \emph{Species}  & 1  &  00:00:00   & 1,176  & 00:00:03  & 67,323   &  00:03:09 & 0          & 00:00:00 \\
  9 \emph{Species}  & 0  &  00:00:00   & 1,073  & 00:00:03  & 223,775  &  00:12:48 & 2,439,310  & 13:31:19 \\
  10 \emph{Species} & 0  &  00:00:00   & 385    & 00:00:02  & 328,397  &  00:19:30 & 4,669,000\textsuperscript{$*$} & 47:39:39 \\
  \bottomrule
\end{tabular}
   \end{center}
  \Caption{Number of enumerated feed-forward, non-competitive CRNs with at least two dual-rail inputs (4 actual species) and two outputs (4 actual species).
    Star (\textsuperscript{$*$}) denotes that the scope has been partially enumerated.}
  \label{tb:2input2outputCount}
\EndTable

\textbf{Smallest \emph{minmax} CRN.}
\emph{Minmax} CRN accepts two inputs and has two outputs,
where one output computes \emph{max}, and other output computes \emph{min} of the inputs.
Since species are in dual-rail form, there is $4$ input and $4$ output species.
Thus, for \emph{minmax} search we enumerated CRNs that have at least $8$ species,
where at least $4$ species only appear as products (output species candidates),
and at least $4$ species which do not appear only as products (input species candidates).
We have further restricted the CRNs to have a total of at most $16$ reactants and products over all reactions.
Enumeration results with those constraints are shown in Table~\ref{tb:2input2outputCount} (isomorphic breaking is imperfect in this case as well).
We discovered the minimal \emph{minmax} CRN,
which is shown in \figname~\ref{fig:min-crns}.
We performed several optimizations to speed up the analysis phase which are described in Appendix~\ref{app:optimization}.

\BeginTable
  \begin{center}
\begin{tabular}{c|rr|rr|rr|rr|rr}
  \toprule
  & \TNC{2}{1 \emph{Reaction}}  & \TNC{2}{2 \emph{Reactions}} & \TNC{2}{3 \emph{Reactions}} & \TNC{2}{4 \emph{Reactions}} & \TNC{2}{5 \emph{Reactions}} \\
  \midrule
  1 \emph{Domain} & \UseMacro{seesaw_D1_R1_S20_crns} & \UseMacro{seesaw_D1_R1_S20_time} & \UseMacro{seesaw_D1_R2_S20_crns} & \UseMacro{seesaw_D1_R2_S20_time} & \UseMacro{seesaw_D1_R3_S20_crns} & \UseMacro{seesaw_D1_R3_S20_time} & \UseMacro{seesaw_D1_R4_S20_crns} & \UseMacro{seesaw_D1_R4_S20_time} & \UseMacro{seesaw_D1_R5_S20_crns} & \UseMacro{seesaw_D1_R5_S20_time} \\
  2 \emph{Domains} & \UseMacro{seesaw_D2_R1_S20_crns} & \UseMacro{seesaw_D2_R1_S20_time} & \UseMacro{seesaw_D2_R2_S20_crns} & \UseMacro{seesaw_D2_R2_S20_time} & \UseMacro{seesaw_D2_R3_S20_crns} & \UseMacro{seesaw_D2_R3_S20_time} & \UseMacro{seesaw_D2_R4_S20_crns} & \UseMacro{seesaw_D2_R4_S20_time} & \UseMacro{seesaw_D2_R5_S20_crns} & \UseMacro{seesaw_D2_R5_S20_time} \\
  3 \emph{Domains} & \UseMacro{seesaw_D3_R1_S20_crns} & \UseMacro{seesaw_D3_R1_S20_time} & \UseMacro{seesaw_D3_R2_S20_crns} & \UseMacro{seesaw_D3_R2_S20_time} & \UseMacro{seesaw_D3_R3_S20_crns} & \UseMacro{seesaw_D3_R3_S20_time} & \UseMacro{seesaw_D3_R4_S20_crns} & \UseMacro{seesaw_D3_R4_S20_time} & \UseMacro{seesaw_D3_R5_S20_crns} & \UseMacro{seesaw_D3_R5_S20_time} \\
  4 \emph{Domains} & \UseMacro{seesaw_D4_R1_S20_crns} & \UseMacro{seesaw_D4_R1_S20_time} & \UseMacro{seesaw_D4_R2_S20_crns} & \UseMacro{seesaw_D4_R2_S20_time} & \UseMacro{seesaw_D4_R3_S20_crns} & \UseMacro{seesaw_D4_R3_S20_time} & \UseMacro{seesaw_D4_R4_S20_crns} & \UseMacro{seesaw_D4_R4_S20_time} & \UseMacro{seesaw_D4_R5_S20_crns} & \UseMacro{seesaw_D4_R5_S20_time} \\
  5 \emph{Domains} & \UseMacro{seesaw_D5_R1_S20_crns} & \UseMacro{seesaw_D5_R1_S20_time} & \UseMacro{seesaw_D5_R2_S20_crns} & \UseMacro{seesaw_D5_R2_S20_time} & \UseMacro{seesaw_D5_R3_S20_crns} & \UseMacro{seesaw_D5_R3_S20_time} & \UseMacro{seesaw_D5_R4_S20_crns} & \UseMacro{seesaw_D5_R4_S20_time} & \UseMacro{seesaw_D5_R5_S20_crns} & \UseMacro{seesaw_D5_R5_S20_time} \\
  6 \emph{Domains} & \UseMacro{seesaw_D6_R1_S20_crns} & \UseMacro{seesaw_D6_R1_S20_time} & \UseMacro{seesaw_D6_R2_S20_crns} & \UseMacro{seesaw_D6_R2_S20_time} & \UseMacro{seesaw_D6_R3_S20_crns} & \UseMacro{seesaw_D6_R3_S20_time} & \UseMacro{seesaw_D6_R4_S20_crns} & \UseMacro{seesaw_D6_R4_S20_time} & \UseMacro{seesaw_D6_R5_S20_crns} & \UseMacro{seesaw_D6_R5_S20_time} \\
  \bottomrule
\end{tabular}
   \end{center}
  \Caption{Number of enumerated seesaw reactions with different number of domains and reactions, and up to $20$ distinct species.}
  \label{tb:seesaw-enum}
  \vspace{-15pt}
\EndTable  

\textbf{Seesaw enumeration.}
We enumerated all nonisomorphic seesaw CRNs up to specified bounds on the number of domains and reactions.
Table~\ref{tb:seesaw-enum} shows the number of enumerated CRNs restricted to $1$-$5$ reactions, $1$-$6$ domains, and up to $20$ species.
Since $5$ seesaw reactions can have at most $20$ distinct species this includes all possible seesaw CRNs in the scope of $1$-$5$ reactions.
For seesaw networks, we define isomorphic CRNs as those that can be obtained by:
(a) swapping domain names,
(b) changing order of reactants or products,
(c) changing order of reactions,
(d) swapping reactants with products (follows from the reversibility of seesaw reactions).

In order to check for isomorphisms while enumerating seesaw CRNs, 
we maintain a set of previously enumerated CRNs and all their isomorphisms. 
If a newly enumerated CRN is not found in the current set, 
we create the isomorphic class of the CRN by making all permutations of the CRN, and adding them to the set.
Permutations are done only with respect to domains.
Permuting the order of reactants and products, as well as swapping reactants and products, is not needed as we follow the convention of enumerating CRNs in a form $S_{??} + LG_{??} \leftrightarrow S_{??} + RG_{??}$.
Permuting the order of reactions is not needed, as the set of CRNs is preserved as a hash table where a custom-made hash function is used for CRNs (a same hash value is returned for a CRN irrespective of the order of reactions).
The isomorphic breaking is implemented as a post-processing step in Java.
The run-times reported in Table~\ref{tb:seesaw-enum} include both generation and isomorphic breaking times.

Note that we require that the CRN corresponding to a seesaw system contain all reactions that can occur.
For illustration, we analyze seesaw CRNs with $2$ domains and $1$ reaction.
Due to the reversibility of seesaw reactions we can limit our analysis to CRNs that have a left gate on the left hand side; thus our CRN will be of the form
$S_{??} + LG_{??} \leftrightarrow S_{??} + RG_{??}$, where ? represent domains to be filled in.
We denote two available domains with $a$ and $b$, and we enforce that both domains are used in a CRN.
The possible combinations for the domains of the first strand are $\{aa, ab, ba, bb\}$, where we can remove cases starting with $b$ as they are symmetrical.
Choosing $S_{aa}$ as a first strand, the only option for left gate is $LG_{ab}$ as we have to use two domains and left domain of $LG$ must match right domain of $S$.
This leads to a CRN: $S_{aa} + LG_{ab} \leftrightarrow S_{ab} + RG_{aa}$.
Note that this CRN is not a valid one, as in this case $S_{aa}$ and $RG_{aa}$ can also interact creating additional reaction.
Another option for the strand is $S_{ab}$, in which case there are two options for left gate $LG_{bb}$ and $LG_{ba}$.
In a case of $LG_{bb}$ reaction is following: $S_{ab} + LG_{bb} \leftrightarrow S_{bb} + RG_{ab}$.
This is also not a valid CRN since $S_{bb}$ and $LG_{bb}$ can interact creating additional reaction.
The final option is $S_{ab} + LG_{ba} \leftrightarrow S_{ba} + RG_{ab}$, which is only valid seesaw CRN in a case of $2$ domains and $1$ reaction;
thus Table~\ref{tb:seesaw-enum} shows count $1$ for seesaw CRNs with $2$ domains and $1$ reaction.

Similarly, note that there are $0$ CRNs with $2$ domains and $3$ reactions, but there are $2$ with $2$ domains and $4$ reactions.
This is due to the fact that all $3$ reaction CRNs with $2$ domains have some other species that can also interact producing additional (spurious) reaction.
A curious reader can check that removing any reaction from $4$ reaction $2$ domain seesaw CRNs (Table~\ref{eq:seesaw4rxns}) will leave some species that can interact creating the fourth reaction.

\begin{small}
\begin{table}[!htbp] %
\begin{tabular}{ccc}
\begin{minipage}{0.5\textwidth}
\begin{align}
  \ce{$S_{aa}$ + $LG_{aa}$ &<->[] $S_{aa}$ + $RG_{aa}$} \nonumber \\
  \ce{$S_{ba}$ + $LG_{aa}$ &<->[] $S_{aa}$ + $RG_{ba}$} \nonumber \\
  \ce{$S_{bb}$ + $LG_{ba}$ &<->[] $S_{ba}$ + $RG_{bb}$} \nonumber \\
  \ce{$S_{bb}$ + $LG_{bb}$ &<->[] $S_{bb}$ + $RG_{bb}$} \nonumber
\end{align}
\end{minipage}
&
\begin{minipage}{0.5\textwidth}
\begin{align}
  \ce{$S_{ab}$ + $LG_{ba}$ &<->[] $S_{ba}$ + $RG_{ab}$} \nonumber \\
  \ce{$S_{bb}$ + $LG_{ba}$ &<->[] $S_{ba}$ + $RG_{bb}$} \nonumber \\
  \ce{$S_{ab}$ + $LG_{bb}$ &<->[] $S_{bb}$ + $RG_{ab}$} \nonumber \\
  \ce{$S_{bb}$ + $LG_{bb}$ &<->[] $S_{bb}$ + $RG_{bb}$} \nonumber
\end{align}
\end{minipage}
\end{tabular}
\vspace{10pt}
\Caption{Seesaw CRNs with $2$ domains and $4$ reactions.}
\label{eq:seesaw4rxns}
\vspace{-15pt}
\end{table}
\end{small}

\Section{Related Work}
\label{sec:related:work}

\textbf{CRN Enumeration.}
Deckard et al.~\cite{deckard2009enumeration} developed an online library of reaction networks,
which was extended~\cite{banaji2017counting} to
catalog reactions of several classes.
These approaches generate non-isomorphic bipartite graphs (two types of vertices for species and reactions) with undirected edges relying on Nauty library~\cite{McKay201494PracticalGraphIsomorphism}.
Each such constructed graph is then reified as multiple CRN instances.
Recent generalization of this work gives the first complete count of all 2-species bimolecular CRNs, and counts for other classes of CRNs such as mass-conserving and reversible~\cite{spaccasassi2019fast}.
Rather than focusing on removing all isomorphisms and generating exact counts of non-isomorphic CRNs in each class, our work allows the user to flexibly specify and analyze structural properties of CRNs of interest (enabling direct generation of CRNs following the structure).
For example, it is not clear how to encode molecular structure (such as we do for seesaw networks) using graph-based models.

\textbf{Minimal Systems with Desired Behavior.}
Complementary to CRN enumeration, previous work also tackled the problem of finding minimal CRNs respecting some desired properties or exhibiting certain behavior.
Wilhelm~\cite{Wilhelm09SmallestCRNWithBistability} discovers the smallest elementary CRN with bistability.
Wilhelm and Heinrich~\cite{WilhelmHeinrich95SmallestCRNWithHopfBifurcation} similarly detect the smallest CRN with Hopf bifurcation.
In comparison with this line of work, our paper presents a more general framework that allows specifying structure and properties, including different functions, of CRNs to be explored.

Recent work due to Murphy et al~\cite{murphy2018synthesizing} is close to ours in spirit, but focuses on discrete-state stochastic systems (integer molecular counts of the species), rate-dependent reactions, and does not guarantee that discovered CRNs are minimal.
Cardelli et al~\cite{cardelli2017syntax} take a program synthesis approach to generate CRNs that follow properties provided by a certain ``sketch'' language (i.e., a template) using SMT solvers on the back end~\cite{de2008z3,Barrett2011CVC20323052032319}.

\textbf{Computational power of CRNs.}
Much ongoing work has explored computational power of CRNs~\cite{SalehiETAL17CRNsForComputingPolynomials,Magnasco97ChemicalKineticsIsTuringUniversal,HuangETAL12CompilingControlFlowIntoBiochemicalReactions,vasic2018crn++}.
It is shown how to map complex computation to CRNs, such as mapping polynomials to chemical reactions,
mapping discrete algorithms, and even defining a high-level imperative languages that map to CRNs.
We believe that by exploring CRNs bottom up, we may found answers of what the appropriate (more efficient) high-level primitives are to be used for implementing such high-level functionality.

\Section{Conclusion}

We introduced the use of Alloy, a framework for modeling and analyzing structural constraints and behavior in software systems, 
to enumerate CRNs with declaratively specified properties. 
We showed how this framework can enumerate CRNs with a variety of structural constraints 
including biologically motivated catalytic networks and metabolic networks, 
and seesaw networks motivated by DNA nanotechnology. 
We also used the framework to explore analog function computation in rate-independent CRNs. 
We applied our approach in a case-study to find the smallest CRNs computing the \emph{max}, \emph{minmax}, \emph{abs} and \emph{ReLU} functions in a natural subclass of rate-independent CRNs where rate-independence follows from structural network properties.

There remain a number of open questions that motivate future research directions.
An important area of optimization is improving the run-time of the Alloy enumeration. 
Can we optimize the isomorphic breaking process to  eliminate all isomorphisms?
For improved efficiency and ease of use, do we need to rely on a separate tool like Mathematica to determine whether a given CRN computes the desired function, or can the necessary functionality be performed in Alloy alone?
Finally, it remains to be seen how easily the techniques developed in this paper could be applied to rate-dependent computation.

\bibliography{bib}

\newpage
\appendix

\section{Proof of Rate Independence}
\label{app:rateindependence}

In this section we develop an argument that the class of feed-forward, non-competitive CRNs as defined in the main text is rate-independent.
For simplicity, we base our argument on the discrete CRN model, in which concentrations are integer molecular counts, reactions are discrete events (firings), and rate-independence corresponds to behaving correctly no matter what order the reactions occur in~\cite{chen2014deterministic}.
The continuous model is usually taken as an approximation of the discrete model.

Note that when we say that a species $S$ is consumed by a reaction, we mean that it appears with negative net stoichiometry in the reaction.
So we would not say that a catalyst is consumed.
We define produced similarly.
We say configuration $d$ is reachable from $c$ if there is a sequence of reactions that can fire to get from $c$ to $d$.

In the main text, we define non-competitive as follows: if a species is consumed in a reaction then it cannot appear as a reactant somewhere else.
Feed-forward is defined as follows: there exists a total ordering on the reactions such that no reaction consumes a species produced by a reaction later in the ordering.
We also require that all reactions consume some species (boundedness condition).

Here we show that the feed-forward condition combined with boundedness implies that the CRN will always reach a static equilibrium.
(A static equilibrium is one where no reaction can fire.)
We then show that adding the non-competitive condition implies that the CRN always reaches the same static equilibrium independent of the order in which the reactions happen to occur.

\emph{The CRN always reaches some static equilibrium}:
If not then there is a set of reactions that can fire infinitely often.
Choose the earliest (according to the ordering) reaction in this set.
It must consume some $S$ by boundedness.
But by feed-forwardness, $S$ can only be produced earlier in the ordering.
Which means that the reactions that net produce $S$ can only fire finite many times (they are not in this set). This is a contradiction.

\emph{The CRN always reaches the same static equilibrium}:
Toward a contradiction, suppose two different static equilibria $c$ and $d$ are reachable.
Let $p$ be the path to $c$ and $q$ be the path to $d$.
Without loss of generality there are reactions that fire fewer times in $p$ than in $q$.
Let $R$ be the reaction among these that comes earliest in the ordering.
So compared to $q$, $p$ has at least as many firings of reactions earlier in the ordering than $R$.
By non-competitiveness, no other reaction consumes the reactants of $R$.
Let $S$ be a reactant of $R$.
Consider two cases:
(1) $S$ is consumed in $R$.
By feed-forwardness, $S$ must be produced in a reaction earlier in the ordering than $R$.
This means that the reactions producing $S$ fire at least as much in $p$ as in $q$.
Since $R$ fired fewer times in $p$ than in $q$, there are some of $S$ left in $c$.
(2) $S$ is not consumed in $R$ (it acts as a catalyst).
By the argument below, since $R$ fires in $q$ at least once, $R$ fires in $p$ at least once. 
Thus $S$ is present in $c$. 
Combining (1) and (2), we have that $R$ can fire in $c$, which contradicts the assumption that $c$ is a static equilibrium.

\emph{There are no reactions that can fire on the path toward one static equilibrium but not fire on the path to another}:
Toward a contradiction, suppose two different static equilibria $c$ and $d$ are reachable.
Let $p$ be the path to $c$ and $q$ be the path to $d$.
Let $\Omega$ be the set of reactions that fire in $q$ but not in $p$. 
Let $R$ be the reaction in $\Omega$ that occurs first (in time) in $q$.
Its reactants must be either inputs or produced outside of $\Omega$ since $R$ is the first reaction in $\Omega$ that fired in $q$.
By non-competitiveness, the reactants of $R$ cannot be consumed in any reaction other than $R$.
So it must be possible to fire $R$ at the end of $p$, which contradicts the assumption that $p$ is a static equilibrium. 

\section{Background: Alloy}
\label{app:alloy}

The Alloy modeling language is a first-order logic with transitive
closure~\cite{Jackson02Alloy}.  The Alloy analyzer is a fully
automatic tool for \emph{scope-bounded} analysis of properties of
Alloy models~\cite{JacksonETAL00ALCOA}.  Given an Alloy model and a
\emph{scope}, i.e., a bound on the universe of discourse, the analyzer
translates the Alloy model to a propositional satisfiability (SAT)
formula and invokes an off-the-shelf SAT
solver~\cite{EenSorensson03MiniSAT} to analyze the model.

An Alloy model consists of a set of paragraphs where each paragraph
declares some typed sets or relations, defines some logical
constraints, or defines a command that informs the analyzer of the
analysis to perform.  Each command defines a constraint solving
problem. and each solution to the problem defines an Alloy
\emph{instance}, i.e., a valuation of the sets and relations declared
in the model such that the constraints with respect to the command are
satisfied.  The analyzer supports instance enumeration using
incremental SAT
solvers~\cite{MoskewiczETAL01Chaff,EenSorensson03MiniSAT}.  In
addition, the analyzer supports \emph{symmetry breaking} and adds
symmetry breaking
predicates~\cite{Shlyakhter01EffectiveSymmetryBreaking} to the
original formula, which allows the backend SAT solvers to more
effectively prune their search, and when enumerating solutions, create
fewer solutions~\cite{KhurshidETAL03Enumeration}.  The analyzer's
default symmetry breaking does not guarantee removal of all
isomorphisms but is quite effective in practice.

\section{Autocatalytic Reactions}
\label{app:autocatalytic}

\begin{figure}[!t]
\begin{CodeOut}
\begin{verbatim}
module autocatalytic
open elementary
pred Autocatalytic[] { Elementary[] and all r: Reaction | AutocatalyticReaction[r] }
pred AutocatalyticReaction[r: Reaction] {
    some elems[r.reactants] & elems[r.products]
    eq[#r.products, 2] and eq[#elems[r.products], 1] }
\end{verbatim}
\end{CodeOut}
\caption{Autocatalytic reactions.\label{fig:autocatalytic}}
\end{figure}

Similarly to catalytic reactions we model autocatalytic (\figname~\ref{fig:autocatalytic}).
Autocatalytic reactions add a requirement that in addition to existence of a catalyst species,
the catalyst converts the other species into itself, for example: $ \ce{$X + Y \to Y + Y$} $.
\section{ReLU Minimality}
\label{app:reluminimality}

In this section we argue that our enumeration in Table~\ref{tb:enumcount} is sufficient to ensure that 5 species are necessary for computing \Relu no matter how many reactions are allowed.

Because with $4$ species there are at most $2$ different reactions possible (which we enumerate).
Consider the \Relu CRN with $4$ species.
This CRN must consist of $2$ input species ($X^+$ and $X^-$) and $2$ output species ($Y^+$ and $Y^-$), which we require to be distinct.
Further, the output species have to appear only as products.
Thus, only species $X^+$ and $X^-$ can appear as reactants.
Due to the requirement that every reaction has to net consume some species (\figname~\ref{fig:feedforward}),
and that different reactions have to consume different species (non-competitiveness),
it follows that the CRN can have at maximum $2$ reactions,
one net consuming $X^+$, and other $X^+$ species.
Considering that our technique did not discover any \Relu CRN with $2$ reactions and $4$ species,
we conclude that there is no \Relu computing CRN with $4$ species.
\section{Optimizing Analysis}
\label{app:optimization}

In this section we explain how we optimize the analysis phase of search for \emph{minmax} CRN.

The optimization is done by including tests.
Instead of invoking \textit{FindInstance} SMT solver for every combination of inputs and outputs,
we construct a set of concrete test cases.
If a test case fails we immediately discard that combination and move to the next one.
This optimization improved analysis from $75$s to $7.3$s measured on the discovered \emph{minmax} CRN.
Furthermore from equality $|max(a,b)|+|min(a,b)| = min(|a|,|b|) + max(|a|,|b|)$,
we first checked for CRNs that sattisfy this condition (using tests and FindInstance),
and only run the check whether output species compute min and max on those.
Checking for the above equality speeded up analysis becase the equality does not depend on the order of output species $y1$ and $y2$, thus reducing number of input output combinations that need to be tried.
After implementing this additional optimization step analysis time went down to $0.75$s measured on the discovered \emph{minmax} CRN.
The optimizations made it feasible to discover the \emph{minmax} CRN.
\section{Symmetry breaking}
\label{app:symmetry}

This section shows our Alloy model for symmetry breaking of CRNs
(\figname~\ref{fig:symmetry}).

The Alloy analyzer during its translation from Alloy to propositional
formulas automatically adds to the propositional formulas
\emph{symmetry breaking} predicates, which reduce the number of
isomorphic solutions~\cite{Shlyakhter01EffectiveSymmetryBreaking}.
However, this automatic support is not practical for breaking all
isomorphisms since there is a delicate trade-off between the
complexity of the predicates that are added and the time it takes for
the back-end solvers to handle them.

We follow a more effective approach where additional constraints
\emph{in Alloy} are mechanically added directly to the Alloy
model~\cite{KhurshidETAL03Enumeration}.  The key idea is to define a
linear order on the atoms and require that any solution when scanned
in a pre-defined manner contains the atoms in conformance with the
linear order.  The approach breaks all symmetries for rooted,
edge-labeled graphs.  However, CRNs represent a more complex structure
and the approach does not guarantee breaking all symmetries.
Nonetheless, it removes many isomorphic solutions and provides us a
practical tool for exploring CRNs.

Note that the symmetry breaking is focused on a case of elementary CRNs as those CRNs are our focus group
(all of our inherited CRN models are subclass of elementary).
\XComment{First, the ordering of species and reactions is created using the Alloy utility module \textit{ordering}.
Predicate \CodeIn{CheckFirstReaction}
TODOOOO.....
}

\begin{figure}[!t]
\begin{CodeOut}
\begin{verbatim}
module symmetry

open elementary

open util/ordering[Species] as Sordering
open util/ordering[Reaction] as Rordering

pred CheckFirstReaction {
  let first = Rordering/first, 
      r1 = 0.(first.reactants), r2 = 1.(first.reactants),
      p1 = 0.(first.products), p2 = 1.(first.products)
  {
    r1 = Sordering/first
    r2 in r1 + r1.next
    p1 in r1 + r2 + (r1 + r2).next
    p2 in r1 + r2 + p1 + (r1 + r2 + p1).next
  }
}

pred CheckNonFirstReaction() {
  all r: Reaction - Rordering/first {
    let prevRxns = Rordering/prevs[r],
        prevSpecies = Int.(prevRxns.reactants + prevRxns.products),
        r1 = 0.(r.reactants), r2 = 1.(r.reactants),
        p1 = 0.(r.products), p2 = 1.(r.products)
    {
      r1 in prevSpecies + prevSpecies.next
      r2 in prevSpecies + r1 + (prevSpecies + r1).next
      p1 in prevSpecies + r1 + r2 + (prevSpecies + r1 + r2).next
      p2 in prevSpecies + r1 + r2 + p1 + (prevSpecies + r1 + r2 + p1).next
    }
  }
}

pred OrderReactionsBySize() {
  all disj r1, r2 : Reaction {
    Rordering/lt[r1, r2] implies {
      lt[#r1.reactants, #r2.reactants]
      or (eq[#r1.reactants, #r2.reactants] 
            and lte[#r1.products, #r2.products])
    }
  }
}

pred ReactionsSameSize[r1, r2: Reaction] {
  eq[#r1.reactants, #r2.reactants]
    and eq[#r1.products, #r2.products]
}

pred CheckLexicographic() {
  all r: Reaction - Rordering/first {
    let p = r.prev,
    rr1 = 0.(r.reactants), rr2 = 1.(r.reactants), rp1 = 0.(r.products), rp2 = 1.(r.products),
    pr1 = 0.(p.reactants), pr2 = 1.(p.reactants), pp1 = 0.(p.products), pp2 = 1.(p.products)
    {
       ReactionsSameSize[r, p] implies {
        // DO only if sizes are the same assuming the size constraing.
        rr1 in pr1.*next
        rr1 = pr1 implies (no pr2 or rr2 in pr2.*next)
        (rr1 = pr1 and rr2 = pr2) implies (rp1 in pp1.*next)
        (rr1 = pr1 and rr2 = pr2 and rp1 = pp1) implies (no pp2 or rp2 in pp2.*next)
       }
    }
  }

  all r: Reaction {
    let r1 = 0.(r.reactants), r2 = 1.(r.reactants), p1 = 0.(r.products), p2 = 1.(r.products)
    {
      some r1 and some r2 implies Sordering/lte[r1, r2]
      some p1 and some p2 implies Sordering/lte[p1, p2]
    }
  }
}

pred SymmetryBreaking {
  Elementary
  CheckFirstReaction
  CheckNonFirstReaction
  OrderReactionsBySize
  CheckLexicographic
}
\end{verbatim}
\end{CodeOut}
\caption{Alloy modeling of CRN symmetry breaking.}
\label{fig:symmetry}
\end{figure}

\end{document}